\documentclass{article}

\usepackage{amsmath}
\usepackage{graphicx}
\usepackage{hyperref}
\usepackage{url}
\usepackage[colorinlistoftodos, textwidth=1.5\marginparwidth, shadow]{todonotes}

\newcommand{\julia}{\text{Julia}}

\newcommand{\code}[1]{\texttt{#1}}

\newlength{\boxwidth}
     \renewcommand{\,}{\hspace*{1em}   }

          \renewcommand{\>}{\hspace*{0.1em}   }
    \usepackage{graphicx}
\usepackage{color}
\usepackage{framed}

\definecolor{shadecolor}{gray}{.92}
\definecolor{incolor}{rgb}{0,0,.7}
\definecolor{outcolor}{rgb}{.65,0,0}
\definecolor{syntaxcolor}{rgb}{.65,0,0}
\usepackage{verbatim}
\newcommand{\sh}[1]{\textcolor{syntaxcolor}{#1}}

\newcounter{jcounter}
\newenvironment{jinput}[1][]{\ifx#1\relax\else\setcounter{jcounter}{#1}\addtocounter{jcounter}{-1}\fi\refstepcounter{jcounter}\ttfamily\hspace*{-.7in}\noindent\begin{minipage}[t]{.19\textwidth}\vskip2ex\hspace*{\fill}\textcolor{incolor}{In[\arabic{jcounter}]: }\end{minipage}\begin{minipage}[t]{.8\textwidth}\vskip-0ex\begin{shaded}}{\end{shaded}\end{minipage}\par}

\newenvironment{joutput}[1][]{\vskip1ex plus .2ex minus .1ex\ifx#1\relax\else\setcounter{jcounter}{#1}\fi\addtocounter{jcounter}{-1}\refstepcounter{jcounter}\ttfamily\noindent\hspace*{-.5in}\begin{minipage}[t]{.19\textwidth}\vskip0ex\hspace*{\fill}\textcolor{outcolor}{Out[\arabic{jcounter}]: }\end{minipage}\begin{minipage}[t]{.8\textwidth}\vskip-0ex}{\end{minipage}\par\vskip1.5ex}

     \newcommand{\ja}{\begin{jinput}}
     \newcommand{\jb}{\end{jinput}\begin{joutput}}
     \newcommand{\jc}{\end{joutput}}

    \newcommand{\jav}{\begin{jinput}\begin{verbatim}}
     \newcommand{\jbv}{\end{verbatim}\end{jinput}\begin{joutput}\begin{verbatim}}
     \newcommand{\jcv}{\end{verbatim} \end{joutput}}

\usepackage{fancyvrb}

\begin{document}

\setlength{\pdfpageheight}{\paperheight}
\setlength{\pdfpagewidth}{\paperwidth}

\title{Julia: A fresh approach to numerical computing}
\author{Jeff Bezanson \and Alan Edelman \and Stefan Karpinski \and Viral B. Shah}

\date{MIT and Julia Computing\footnote{\url{http://www.juliacomputing.com}}\\July 7, 2015}


\maketitle

\abstract{

Bridging cultures
that have often been distant, Julia combines expertise from the
diverse fields of computer
science and computational science to create a new approach to
numerical  computing. Julia is  designed to be easy and fast.  Julia
 questions notions generally held as ``laws of nature"  by practitioners of   numerical computing:
\begin{enumerate}
\item  High-level dynamic programs have to be slow,
\item  One must prototype in
one language and then rewrite in another language for speed or
deployment, and
\item There are parts of a system
for the programmer, and other parts best left untouched as they are built by the experts.
\end{enumerate}

We introduce the  Julia programming language and its design ---
a  dance between specialization
and abstraction.
Specialization
allows for custom treatment.
{\it Multiple dispatch},  a  technique from computer science,
picks  the right algorithm for the right circumstance.
Abstraction, what good computation is really about, recognizes what remains the same after
differences are stripped away. Abstractions in mathematics
are captured as code through another technique from computer science,
{\it generic programming}.

 Julia shows that  one can have machine performance without sacrificing human convenience.
}

\newpage

\tableofcontents

\newpage

\section{Scientific computing languages: The Julia innovation}

The original numerical computing language was Fortran, short for
``Formula Translating System'', released in 1957.  Since those
early days, scientists have dreamed of writing high-level, generic
formulas and having them translated automatically into low-level,
efficient machine code, tailored to the particular data types they
need to apply the formulas to.  Fortran made historic strides towards
realization of this dream, and its dominance in so many areas of
high-performance computing is a testament to its remarkable success.

The landscape of computing has changed dramatically over the years.
Modern scientific
computing environments such as Python~\cite{numpy}, R~\cite{Rlang},
Mathematica~\cite{mathematica}, Octave~\cite{Octave},
Matlab~\cite{matlab}, and SciLab~\cite{scilab}, to name some, have grown in popularity
and fall under the general category
known as  \hspace{-.08in}  { {\it dynamic languages} or {\it dynamically typed languages}.
In these programming
languages, programmers write simple, high-level code without any
mention of types like \code{int}, \code{float} or \code{double} that
pervade {\it statically typed languages} such as  C and Fortran.





Many researchers today do their day-to-day
work in dynamic languages.  Still, C and Fortran remain the gold
standard for computationally-intensive problems  for performance.
In as much as the dynamic language programmer has missed out on performance,
the C and Fortran programmer has missed out on productivity.
An unfortunate outcome of the currently popular languages is that the
most challenging areas of numerical computing have benefited the least
from the increased abstraction and productivity offered by higher
level  languages.
The
consequences have been more serious than many
realize.

Julia's innovation is the very combination of productivity and performance.
New users want a quick explanation as to why Julia is fast, and
whether somehow the same ``magic dust" could also be sprinkled on
their traditional scientific computing language.
 Julia is fast because of careful language design and
the right combination of the carefully chosen  technologies that work
very well with each other.
This paper demonstrates some of these technologies using a number of examples.
We  invite the reader to follow along  at  \url{http://juliabox.org} using
Jupyter notebooks \cite{jupyter,jupyternature}  or by downloading Julia \url{http://julialang.org/downloads}.

\subsection{Computing transcends communities}

Numerical computing research has always lived on the boundary of computer science, engineering, mathematics, and computational  sciences.
Readers might enjoy trying to label the ``Top 10 algorithms"\cite{top10} by  field, and may quickly
see that  advances typically transcend any one field  with broader impacts to science
and technology as a whole.

Computing  is more than using an overgrown calculator. It is a cross cutting communication medium.
Research into programming languages  therefore breaks us out of our research boundaries.

The first decade of the 21st century saw a boost in such
research with the High Productivity Computing Systems DARPA funded
projects into the languages such as Chapel \cite{chapelref,chapelpaper}, Fortress \cite{fortressref,fortressspec}
} and  X10 \cite{x10,x10paper}. Also contemporaneous has been a growing acceptance of Python.
Up to around 2009 some of us were working on Star-P, an interactive high performance computing system
for parallelizing various dynamic programming languages. Some excellent resources on Star-P that
discuss these ideas are the Star-P user Guide~\cite{starpug}, the
Star-P Getting Started guide~\cite{starpstart}, and various papers~\cite{starpright,starpstudy,Husbands98inter,Choy04star-p:high}.
Julia  continues our research into parallel computing, with the most
important lesson from our Star-P experience being that one cannot
design a high performance parallel programming system without a
programming language that works well sequentially.


\subsection{Julia architecture and language design philosophy}


Many popular dynamic languages were not designed with the goal of high
performance.  After all, if you wanted really good performance you
would use a static language, or so the popular wisdom would say.  Only with the increased need in the
day-to-day life of scientific programmers for simultaneous
productivity and performance in a single system has the need for
high-performance dynamic languages become pressing.  Unfortunately,
retrofitting an existing slow dynamic language for high performance is
almost impossible \textit{specifically} in numerical computing
ecosystems.  This is because numerical computing requires
performance-critical numerical libraries, which invariably depend on
the details of the internal implementation of the high-level language,
thereby locking in those internal implementation details.
 For example, you can run Python code much faster than the standard
 CPython implementation using the PyPy just-in-time compiler; but PyPy
 is currently  incompatible with NumPy and the rest of SciPy.

Another important point is that just because a program is available in
C or Fortran, it may not run efficiently from the high level language
or be easy to ``glue" it in.  For example when Steven Johnson tried to
include his C \verb+erf+ function in Python, he reported that Pauli
Virtane had to write glue
code\footnote{\url{https://github.com/scipy/scipy/commit/ed14bf0}} to
vectorize the erf function over the native structures in Python in
order to get good performance. Johnson also had to write similar glue
code for Matlab, Octave, and Scilab. The Julia effort was, by
contrast, effortless.\footnote{Steven Johnson, personal
  communication. See
  \url{http://ab-initio.mit.edu/wiki/index.php/Faddeeva_Package}} As
another example, \verb+randn+, Julia's normal random number generator
was originally based on calling \verb+randmtzig+, a C implementation.
It turned out later, that a pure Julia implementation of the same code
actually  ran faster, and is  now the default implementation.
In  some cases, ``glue'' can often lead to poor performance, even
when the underlying libraries being called are high performance.

The best path to a fast, high-level system for scientific and
numerical computing is to make the system fast enough that all of its
libraries can be written in the high-level language in the first
place. The JUMP.jl~\cite{jump} and the Convex.jl~\cite{convexjl} packages are great examples of the
success of this approach---the entire library is written in Julia and
uses many Julia language features described in this paper.

 {\bf The Two Language Problem:} As long as the developers' language
 is harder than the users' language, numerical computing will always
 be hindered. This is an essential part of the design philosophy of Julia: all basic
functionality must be possible to implement in Julia---never force the
programmer to resort to using C or Fortran.
Julia solves the two language problem.
Basic
functionality must be fast: integer arithmetic, for loops, recursion,
floating-point operations, calling C functions, manipulating C-like
structs.  While these are not only important for numerical programs,
without them, you certainly cannot write fast numerical code.
``Vectorization languages'' like Python+NumPy, R, and Matlab hide
their for loops and integer operations, but they are still there, inside the  C
and Fortran, lurking behind the  thin veneer.  Julia removes this
separation entirely, allowing high-level code to ``just write a for
loop'' if that happens to be the best way to solve a problem.

We believe that the Julia programming language fulfills much of the
Fortran dream: automatic translation of formulas into efficient
executable code.  It allows programmers to write clear, high-level,
generic and abstract code that closely resembles mathematical
formulas, as they have grown accustomed to in dynamic systems, yet
produces fast, low-level machine code that has traditionally only been
generated by static languages.

Julia's ability to combine these levels of performance and
productivity in a single language stems from the choice of a number of
features that work well with each other:

\begin{enumerate}
  \item An expressive type system, allowing optional type
  annotations  (Section \ref{sec:types});
  \item Multiple dispatch using these types to select implementations (Section \ref{sec:select});
   \item Metaprogramming for code generation (Section \ref{subsection:macros});
  \item A dataflow type inference algorithm allowing types of
    most expressions to be inferred \cite{jeff:thesis,Bezanson:2012jf};
  \item Aggressive code specialization against run-time types \cite{jeff:thesis,Bezanson:2012jf};
  \item Just-In-Time (JIT) compilation  \cite{jeff:thesis,Bezanson:2012jf} using the  LLVM  compiler framework~\cite{LLVM}, which is also used
    by a number of other compilers such as Clang~\cite{clang} and
    Apple's Swift~\cite{swift}; and
  \item Julia's carefully written libraries that leverage the language
    design, i.e., points 1 through 6 above (Section \ref{sec:lang}).
\end{enumerate}

Points 1, 2, and 3 above are especially
for the human, and the focus of this paper. On details about the parts
that are about language implementation and internals such as in points
4, 5, and 6, we direct the reader to our earlier
work(~\cite{jeff:thesis,Bezanson:2012jf}). Point 7 brings everything together to build high performance computational libraries in Julia.

Although a sophisticated type system is made available to the
programmer, it remains unobtrusive in the sense that one is never
required to specify types, nor are type annotations necessary for
performance.  Type information flows naturally through the program due to dataflow type inference.

In what follows, we describe the benefits of Julia's language design
for numerical computing, allowing programmers to more readily express
themselves and obtain performance at the same time.



\section{A taste of Julia}
\subsection{A brief tour}



\hspace*{.13in}
\begin{jinput}
A = rand(3,3) + eye(3)  \sh {\# Familiar Syntax} \\
inv(A)
\end{jinput}
\begin{joutput}
3x3 Array\{Float64,2\}:\\
\hspace*{.02em} 0.698106  -0.393074  -0.0480912 \\
 -0.223584 \hspace*{.02em}  0.819635  -0.124946  \\
 -0.344861 \hspace*{.02em}  0.134927 \hspace*{.02em}  0.601952
\end{joutput}

The output from the Julia prompt says that $A$ is a two dimensional
matrix of size $3 \times 3$, and contains double precision floating
point numbers.

Indexing of arrays is performed with brackets, and
is 1-based. It is also possible to compute an entire array
expression and then index into it, without assigning the expression to
a variable:

\begin{jinput}
x = A[1,2]   \\
y = (A+2I)[3,3]     \sh{    \# The [3,3] entry of  A+2I}
\end{jinput}

\begin{joutput}
2.601952
\end{joutput}

In Julia, \verb+I+ is a built-in representation of the identity
matrix,
without explicitly forming the identity matrix as is commonly done
using commands such as ``eye." (``eye" ,
a homonym of  ``I",  is used in such languages as Matlab, Octave,
Go's matrix library, Python's Numpy, and Scilab.)

Julia has symmetric tridiagonal matrices as a special type.  For
example, we may define Gil Strang's favorite matrix (the second order
difference matrix) in a way that uses only $O(n)$ memory.

\begin{figure}[h]
  \centering
  \includegraphics[width=3in]{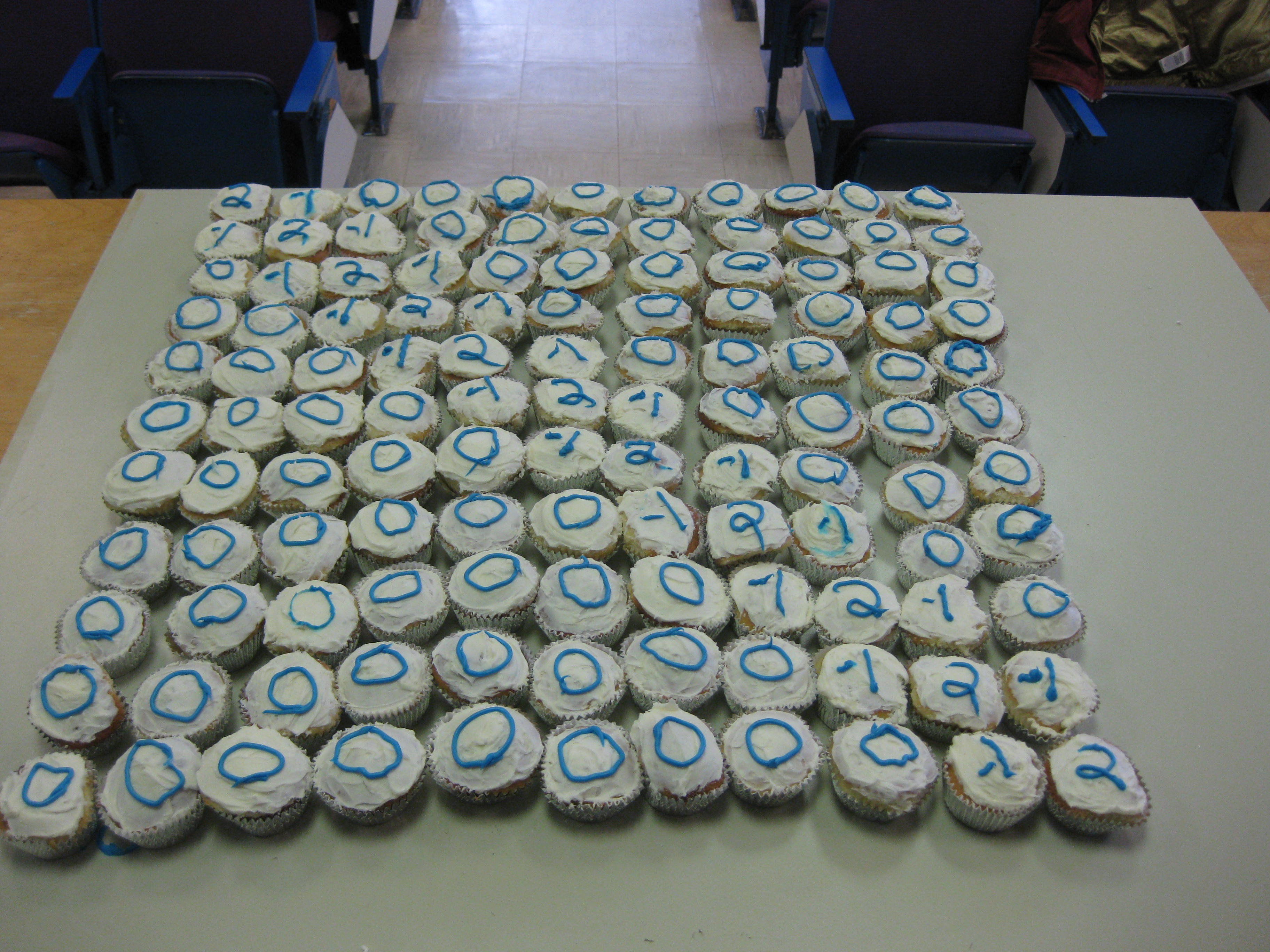}
\caption{Gil Strang's favorite matrix is {\tt strang(n) =
    SymTridiagonal(2*ones(n),-ones(n-1)) } \newline Julia only stores
  the diagonal and off-diagonal.  (Picture taken in Gil Strang's
  classroom.)  }
\end{figure}

\ja
strang(n) = SymTridiagonal(2*ones(n),-ones(n-1)) \\
strang(7)
\jb
7x7 SymTridiagonal\{Float64\}:    \vspace{-.05in}
\begin{verbatim}
  2.0  -1.0   0.0   0.0   0.0   0.0   0.0
 -1.0   2.0  -1.0   0.0   0.0   0.0   0.0
  0.0  -1.0   2.0  -1.0   0.0   0.0   0.0
  0.0   0.0  -1.0   2.0  -1.0   0.0   0.0
  0.0   0.0   0.0  -1.0   2.0  -1.0   0.0
  0.0   0.0   0.0   0.0  -1.0   2.0  -1.0
  0.0   0.0   0.0   0.0   0.0  -1.0   2.0
  \end{verbatim}
\jc

A commonly used notation to express the solution $x$ to the equation $Ax=b$ is
\verb+A\b+.
If  Julia knows that $A$ is a  tridiagonal matrix,
it  uses an efficient  $O(n)$ algorithm:

\ja
strang(8)\textbackslash ones(8)
\jb
8-element Array\{Float64,1\}:   \vspace{-.1in}
\begin{verbatim}
  4.0
  7.0
  9.0
 10.0
 10.0
  9.0
  7.0
  4.0
  \end{verbatim}
\jc

Note the \verb+Array{ElementType,dims}+ syntax.
In the above example, the elements are  64 bit floats or  \verb+Float64+'s.
The \verb+1+ indicates it is a one dimensional vector.

Consider the sorting of complex numbers.   Sometimes it is handy
to have a sort that generalizes the real sort.  This can be done by
sorting first by the real part, and where there are ties, sort by the imaginary
part.  Other times it is handy to use the polar representation, which sorts
by radius then angle.
By default, complex numbers  are incomparable in Julia.

If a numerical computing language ``hard-wires" its sort to be one or the other,
it misses an opportunity.  A sorting algorithm need not depend on details
of what is being compared or how it is being compared.  One can abstract away these
details thereby reusing a sorting algorithm for many different situations.  One can specialize
 later.  Thus alphabetizing strings, sorting
real numbers, or sorting complex numbers in two or more ways all run with the same code.

In Julia, one can turn a complex number \verb+w+ into an ordered pair
of real numbers  (a tuple of length 2) such as the Cartesian form  \verb+(real(w),imag(w))+
or the polar form \verb+(abs(w),angle(w))+.  Tuples are then compared lexicographically
in Julia.  The sort command takes an optional ``less-than" operator, \verb+lt+, which is used to compare
elements when sorting. Note the compact function definition syntax available
in Julia used in the example below and is of the form \verb+f(x,y,...) = <expression>+.

\ja
\sh{\# Cartesian comparison sort of complex numbers} \\
complex\_compare1(w,z) = (real(w),imag(w)) < (real(z),imag(z)) \\
sort([-2,2,-1,im,1], lt = complex\_compare1 )
\jb
5-element Array\{Complex\{Int64\},1\}: \\
 -2+0im \\
 -1+0im \\
 \hspace*{.02em} 0+1im \\
\hspace*{.02em}  1+0im \\
 \hspace*{.02em} 2+0im
\jc

\ja
\sh{\# Polar comparison sort of complex numbers} \\
complex\_compare2(w,z)  = (abs(w),angle(w)) < (abs(z),angle(z)) \\
sort([-2,2,-1,im,1], lt =  complex\_compare2)
\jb
5-element Array\{Complex\{Int64\},1\}: \\
\hspace*{.02em} 1+0im \\
 \hspace*{.02em} 0+1im \\
 -1+0im \\
\hspace*{.02em}  2+0im \\
 -2+0im
\jc

To be sure, experienced computer scientists tend to suspect there is nothing new under the sun.
The C function \verb+qsort()+ takes a \verb+compar+ function.  Nothing really new there.
Python also has custom sorting with a key. Matlab's sort is more basic.
The real contribution of Julia, as will be fleshed out further in this paper, is that
the design of Julia allows custom sorting to be high performance and
flexible and comparable with implementations in other dynamic
languages that are often written in C.


\ja
Pkg.add("PyPlot") \sh{\# Download the PyPlot package} \\
using PyPlot  \sh {\#  load the functionality into Julia} \\
\\
for i=1:5 \\
\,    y=cumsum(randn(500)) \\
\,    plot(y) \\
end \\ \\
 \includegraphics[width=3in]{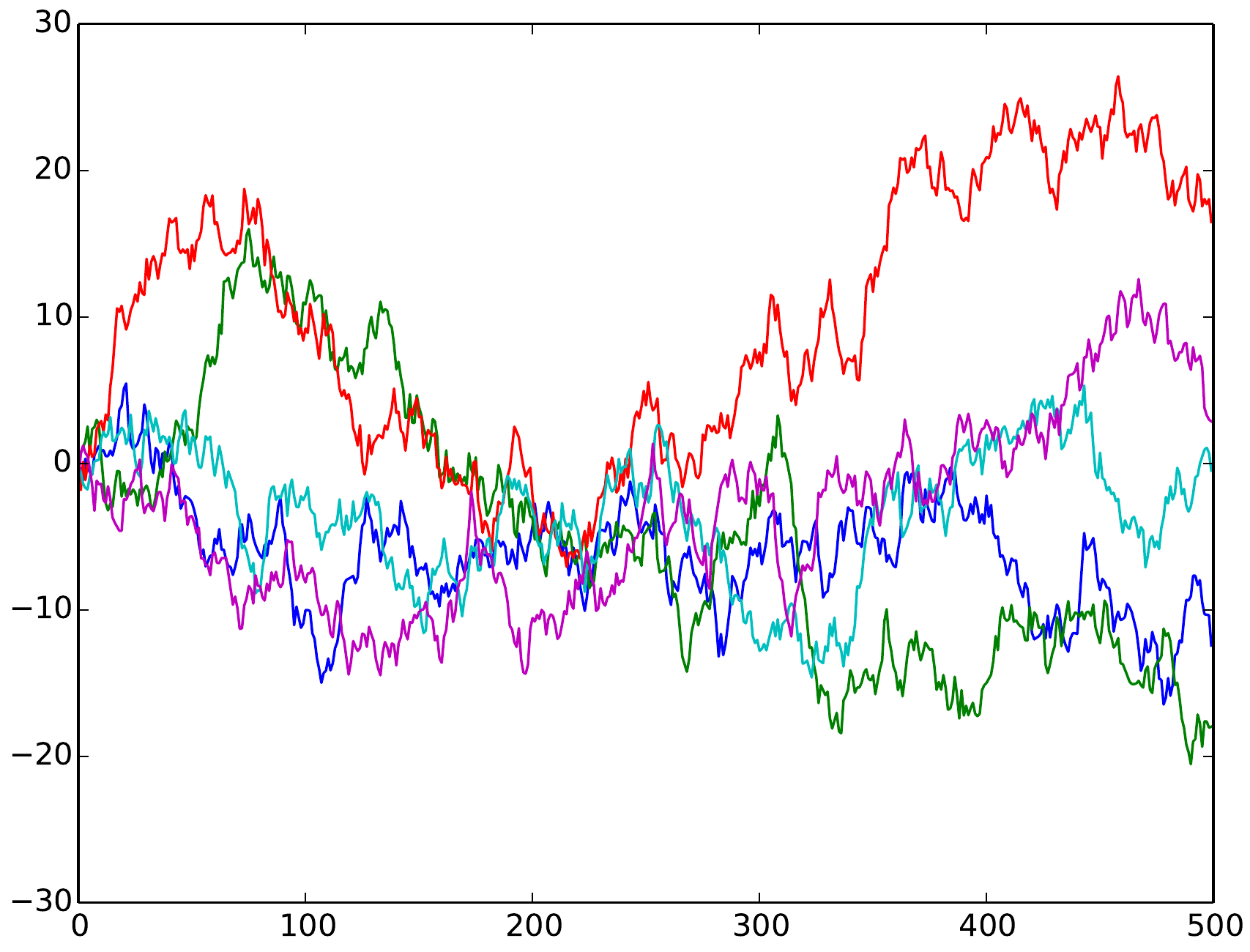}
\end{jinput}

\vspace{.2in}

The next example that we have chosen for the introductory taste of
Julia is a quick plot of Brownian motion, in two ways.
The first such example uses the Python Matplotlib package for graphics, which is popular for users coming from Python or Matlab.
The second example uses Gadfly.jl, another very popular package for plotting.
Gadfly was built by Daniel Jones completely in Julia and was influenced by  the well admired  Grammar of Graphics
(see \cite{gg1} and \cite{gg2})\footnote{See tutorial on \code{http://gadflyjl.org}}
Many Julia users find Gadfly more flexible and
prefer its aesthetics. Julia plots can also be manipulated
interactively with sliders and buttons using Julia's Interact.jl package\footnote{https://github.com/JuliaLang/Interact.jl}.
The Interact.jl package page contains many examples of interactive visualizations\footnote{\code{https://github.com/JuliaLang/Interact.jl/issues/36}}.


\ja
Pkg.add("Gadfly") \sh{\# Download the Gadfly  package} \\
using Gadfly   \sh {\#  load the functionality into Julia} \\
\begin{verbatim}
n = 500
p = [layer(x=1:n, y=cumsum(randn(n)), color=[i], Geom.line)
    for i in ["First","Second","Third"]]
labels=(Guide.xlabel("Time"),Guide.ylabel("Value"),
        Guide.Title("Brownian Motion Trials"),Guide.colorkey("Trial"))
plot(p...,labels...)
\end{verbatim}

\end{jinput}

\vspace{.1in}

\includegraphics[width=4in]{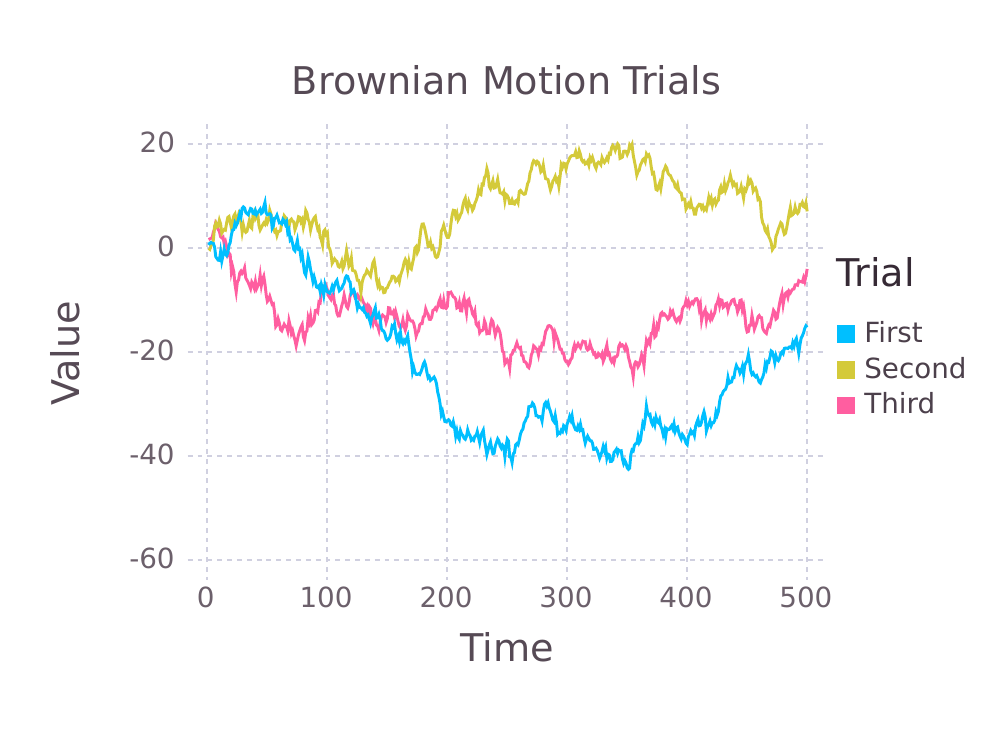}

The ellipses on the last line above is known as a  \verb+splat+  operator.
The elements  of the vector \verb+p+ and the tuple \verb+labels+ are
inserted individually as arguments to the \verb+plot+ function.

%
%
%
%
%


\subsection{An invaluable tool for numerical integrity}

One popular feature of Julia is that it gives the user the ability to
``kick the tires" of a numerical computation.  We thank Velvel Kahan
for the sage advice\footnote{Personal communication, January 2013, in the Kahan
home, Berkeley, California} concerning the importance of this feature.

The idea is simple: a good engineer tests his or her code for numerical stability.
In Julia this can be done by changing IEEE rounding modes.
There are five modes to choose from, yet most engineers silently only choose the
 \verb+RoundNearest+ mode default available in many numerical computing systems.
 If a difference is detected, one can also run the computation in higher precision.
 Kahan writes:

 \begin{quotation}
 Can the effects of roundoff upon a floating-point computation be assessed without submitting it to
a mathematically rigorous and (if feasible at all) time-consuming error-analysis? In general, No.

$\ldots$

 Though far from foolproof, rounding every inexact arithmetic operation (but not
constants) in the same direction for each of two or three directions besides the
default To Nearest is very likely to confirm accidentally exposed hypersensitivity
to roundoff. When feasible, this scheme offers the best {\it Benefit/Cost }ratio.
\cite{kahan:mindless}
\end{quotation}

 As an example, we round a 15x15 Hilbert-like matrix, and take the [1,1] entry of the inverse
 computed in various round off modes.  The radically different answers  dramatically indicates
 the numerical sensitivity to roundoff.
 We even noticed that slight changes to LAPACK give radically different answers.  Very likely
 you will see different numbers when you run this code due to the very high sensitivity to roundoff errors.

\begin{jinput}
h(n)=[1/(i+j+1) for i=1:n,j=1:n]
\end{jinput}
\begin{joutput}
h (generic function with 1 method)
\end{joutput}
\begin{jinput}
 H=h(15); \\
 with\_rounding(Float64,RoundNearest) do \\
    \,    inv(H)[1,1] \\
       end
       \end{jinput}
    \begin{joutput}
154410.55589294434
\end{joutput}
\begin{jinput}
with\_rounding(Float64,RoundUp) do \\
    \,       inv(H)[1,1] \\
       end
       \end{jinput}
       \begin{joutput}
-49499.606132507324
\end{joutput}
\begin{jinput}
 with\_rounding(Float64,RoundDown) do \\
       \,     inv(H)[1,1] \\
       end
       \end{jinput}
       \begin{joutput}
-841819.4371948242
 \end{joutput}

With 300 bits of precision, we obtain \\
\begin{jinput}
with\_bigfloat\_precision(300) do \\
          \,  inv(big(H))[1,1]    \\
           end
           \end{jinput}
           \begin{joutput}
-2.09397179250746270128280174214489516162708857703714959763232689047153\\ 50765882491054998376252e+03
\end{joutput}

Note this is the [1,1] entry of the inverse of the rounded Hilbert-like
matrix, not the inverse of the exact Hilbert-like matrix. Also, the \verb+Float64+  results
are senstive to the BLAS\cite{blas} and LAPACK\cite{lapack}, and may differ on
different machines with different versions of Julia.
For extended precision, Julia uses the MPFR library\cite{MPFR}.  

\subsection{The Julia community}

Julia has been in development since 2009. A public release was
announced in February of 2012.  It is an active open source project
with over 350 contributors and is available under the MIT
License~\cite{mitlicense} for open source software. Over 1.3 million
unique visitors have visited the Julia website since then, and Julia
has now been adopted as a teaching tool in dozens of universities
around the world\footnote{\url{http://julialang.org/community}}.  While it was nurtured at the
Massachusetts Institute of Technology, it is really the contributions
from experts around the world that make it a joy to use for numerical computing.  It is also
recognized as a general purpose computing language unlike
traditional numerical  computing systems, allowing it to be used
not only to prototype numerical algorithms, but also to deploy those
algorithms, and even serve results to the
rest of the world. A great example of this is Shashi Gowda's Escher.jl
package\footnote{https://github.com/shashi/Escher.jl}, which makes it
possible for Julia programmers to build beautiful interactive websites
in Julia, and serve up the results of a Julia computation from the
web server, without any knowledge of HTML or javascript. Another such example is the Sudoku as a service\footnote{http://iaindunning.com/2013/sudoku-as-a-service.html}, by Iain Dunning,
where a Sudoku puzzle is solved using the
  optimization capabilities of the JUMP.jl Julia package \cite{jump} and made
  available as a web service. This is exactly why Julia is being increasingly deployed in production environments in businesses, as seen in various talks at JuliaCon~\footnote{\url{http://www.juliacon.org}}. These use cases utilize not just Julia's capabilities for mathematical computation, but also to build web APIs, database access, and much more. Perhaps most significantly, a rapidly
growing ecosystem of over 600 open source, composable packages
\footnote{url{http://pkg.julialang.org}}, which include a mix of libraries for mathematical as well as general purpose computing, is leading to the adoption of Julia in research, businesses, and in government.

\section{Writing programs with and without types}
\label{sec:types}

\subsection{The balance  between human and the computer}
\label{sec:humancomputer}

Graydon Hoare, author of the Rust programming
language~\cite{rust}, in an essay on ``Interactive Scientific
Computing''~\cite{hoareessay} defined programming languages
succinctly:

\begin{quote}
  Programming languages are mediating devices, interfaces that try to
  strike a balance between human needs and computer needs. Implicit in
  that is the assumption that human and computer needs are equally
  important, or need mediating.
\end{quote}

A program consists of data and operations on data. Data is not just
the input file, but everything that is held---an array, a list, a
graph, a constant---during the life of the program. The more the
computer knows about this data, the better it is at executing
operations on that data. Types are exactly this metadata. Describing
this metadata, the types, takes real effort for the human. Statically
typed languages such as C and Fortran are at one extreme, where all
types must be defined and are statically checked during the
compilation phase. The result is excellent performance. Dynamically
typed languages dispense with type definitions, which leads to greater
productivity, but lower performance as the compiler and the runtime
cannot benefit from the type information that is essential to produce
fast code. Can we strike a balance between the human's preference to
avoid types and the computer's need to know?

\subsection{Julia's recognizable types}

Many users of Julia may never need to know about types for performance.
Julia's type inference system often does the work, giving performance
without type declarations.

Julia's design allows for the gradual learning of concepts, where users
start in a manner that is familiar to them and over time, learn to
structure programs in the ``Julian way'' --- a term that captures
well-structured readable high performance Julia code. Julia users
coming from other numerical computing environments have a notion that
data may be represented as matrices that may be dense, sparse,
symmetric, triangular, or of some other kind. They may also, though
not always, know that elements in these data structures may be single
precision floating point numbers, double precision, or integers of a
specific width. In more general cases, the elements within data
structures may be other data structures. We introduce Julia's type
system using matrices and their number types:

\begin{jinput}
rand(1,2,1)
\end{jinput}
\begin{joutput}
1x2x1 Array\{Float64,3\}: \\
{}[ :, :,  1{}] = \\
\>  0.789166  0.652002
 \end{joutput}


\begin{jinput}
{}[1 2; 3 4{}]
\end{jinput}
\begin{joutput}
2x2 Array\{Int64,2\}: \\
\> 1  2 \\
\>  3  4
 \end{joutput}

\begin{jinput}
{}[true; false{}]
\end{jinput}
\begin{joutput}
2-element Array\{Bool,1\}: \\
\>  true \\
\>  false
 \end{joutput}

 We see a pattern in the examples above. \noindent
 \verb+Array{T,ndims}+ is the general form of the type of a dense
 array with \verb+ndims+ dimensions, whose elements themselves have
a specific type \verb+T+, which is of type double precision floating
point in the first example, a 64-bit signed integer in the second, and
a boolean in the third example.
Therefore \verb+Array{T,1}+ is a 1-d vector (first class
objects in Julia) with element type \verb+T+ and \verb+Array{T,2}+ is
the type for 2-d matrices.

It is useful to think of arrays as a
generic N-d object that may contain elements of any type
\verb+T+. Thus \verb+T+ is a type parameter for an array that can take
on many different values. Similarly, the dimensionality of the array
\verb+ndims+ is also a parameter for the array type. This generality
makes it possible to create arrays of arrays. For example, Using
Julia's array comprehension syntax, we create a 2-element vector
containing $2\times2$ identity matrices.

\begin{jinput}
 a = {}[eye(2) for i=1:2{}]
\end{jinput}
\begin{joutput}
2-element Array\{Array\{Float64,2\},1\}:
\end{joutput}
\noindent

\subsection{User's own types are first class too}
\label{sec:firstclass}


Many dynamic languages for numerical computing have traditionally had
an asymmetry, where built-in types have much higher
performance than any user-defined types. This is not the case with
Julia, where there is no meaningful distinction between user-defined
and ``built-in'' types.

We have mentioned so far a few number types and two matrix types,
\verb+Array{T,2}+ the dense array, with element type \verb+T+ and
\verb+SymTridiagonal{T}+, the symmetric tridiagonal with element type \verb+T+. There
are also other matrix types, for other structures including
SparseMatrixCSC (Compressed Sparse Columns), Hermitian, Triangular, Bidiagonal, and Diagonal. Julia's sparse matrix type
has an added flexibility that it can go beyond storing just numbers as
nonzeros, and instead store any other Julia type as well. The indices
in SparseMatrixCSC can also be represented as integers of any width
(16-bit, 32-bit or 64-bit). All these different matrix types, although
available as built-in types to a user downloading Julia, are
implemented completely in Julia, and are in no way any more or less
special than any other types one may define in their own program.

For demonstration, we create a symmetric arrow matrix type that
contains a diagonal and the first row \verb+A[1,2:n]+.

\ja
\sh{ \# Type Parameter Example (Parameter T)}  \\
\sh{\# Define a Symmetric Arrow Matrix Type with elements of type T} \\

type  SymArrow\{T\}   \\
 \,          dv::Vector\{T\}                     \sh{   \# diagonal  }\\
 \,          ev::Vector\{T\}                     \sh{   \# 1st row{}[2:n{}] }\\
    end \\ \\
    \sh{ \# Create your first Symmetric Arrow Matrix}  \\
S = SymArrow([1,2,3,4,5],[6,7,8,9])
\jb
\begin{verbatim}
SymArrow{Int64}([1,2,3,4,5],[6,7,8,9])
\end{verbatim}
\jc

 The parameter in the array refers to the type of each element of the array.
 Code can and should be written independently of the type of each element.

 Later in Section \ref{sec:arrow}, we develop the symmetric arrow example much further.
 The \verb+SymArrow+
matrix type contains two vectors, one each for the diagonal and the
first row, and these vector contain elements of type \verb+T+. In the type definition,
the type \verb+SymArrow+ is parametrized by the type of the
storage element \verb+T+. By doing so, we have created a generic type,
which refers to a universe of all arrow matrices containing elements
of all types. The matrix \verb+S+, is an example where \verb+T+ is \verb+Int64+.
When we write functions in Section~\ref{sec:arrow}
that operate on arrow matrices, those functions themselves will be
generic and applicable to the entire universe of arrow matrices we
have defined here.

Julia's type system allows for abstract types, concrete ``bits''
types, composite types, and immutable composite types. All of these
can have parameters and users may even write programs using unions of
these different types. We refer the reader to read all about Julia's
type system in the types chapter in the Julia manual\footnote{See the chapter
  on types in the Julia manual:
  \url{http://docs.julialang.org/en/latest/manual/types/}}.

\subsection{Vectorization: Key Strengths and Serious Weaknesses}

Users of traditional  high level computing languages know that vectorization
improves performance.  Do most users know exactly why vectorization
is so useful?  It is precisely because, by vectorizing, the user has promised
the computer that the type of an entire vector of data matches the very first element.
This is an example where users are willing to provide type information to the
computer without even knowing exactly that is what they are doing.
Hence, it is an example of a  strategy that balances the computer's needs with the human's.

From the computer's viewpoint, vectorization means that
operations on data happen largely in sections of the code where types
are known to the runtime system.  When the runtime is operating on
arrays, it has no idea about the data contained in an array until it
encounters the array. Once encountered, the type of the data within
the array is known, and this knowledge is used to execute an
appropriate high performance kernel. Of course what really occurs at
runtime is that the system figures out the type, and gets to reuse
that information through the length of the array.  As long as the
array is not too small, all the extra work in gathering type
information and acting upon it at runtime is amortized over the entire
operation.

The downside of this approach is that the user can achieve
high performance only with built-in types, and user defined types end
up being dramatically slower. The restructuring for vectorization is
often unnatural, and at times not possible. We illustrate this with an
example of the cumulative sum computation. Note that due to the size
of the problem, the computation is memory bound, and one does not observe
the case with complex arithmetic to be twice as slower than the real case,
even though it is performing twice as many floating point operations.

\begin{jinput}
\sh {\# Sum prefix  (cumsum) on vector w  with elements of type T} \\
function prefix\{T\}(w::Vector\{T\})\\
\,    for i=2:size(w,1)\\
\,\,         w[i]+=w[i-1]\\
\,    end\\
\,    w\\
end
\end{jinput}

We execute this code on a vector of double precision numbers and
double-precision complex numbers and observe something
that may seem remarkable: similar running times.

\ja
x = ones(1\_000\_000)\\
@time prefix(x) \\ \\
y = ones(1\_000\_000) + im*ones(1\_000\_000)\\
@time prefix(y);
\jb
elapsed time: 0.003243692 seconds (80 bytes allocated) \\
elapsed time: 0.003290693 seconds (80 bytes allocated)
\jc

\noindent

This simple example is difficult to vectorize, and hence is often
provided as a \verb+built-in+ function in many numerical computing
systems. In Julia, the implementation is very similar to the snippet
of code above, and runs at speeds similar to C. While Julia users can
write vectorized programs like in any other dynamic language,
vectorization is not a pre-requisite for performance. This is because
Julia strikes a different balance between the human and the computer
when it comes to specifying types. Julia allows optional type
annotations, which are essential when writing libraries that utilize
multiple dispatch, but not for end-user programs that are exploring
algorithms or a dataset.

Generally, in Julia, type annotations are not for performance.  They are purely for
code selection. (See Section \ref{sec:select}).
If the
programmer annotates their program with types, the Julia compiler will
use that information. But for the most part,
user code often includes minimal or no type annotations, and the Julia
compiler automatically infers the types.


\subsection{Type inference  rescues ``for loops" and so much  more}
\label{sec:inference}

A key component of Julia's ability to combine performance with
productivity in a single language is its implementation of
dataflow type
inference~\cite{graphfree},\cite{kaplanullman},\cite{Bezanson:2012jf}.
Unlike type inference algorithms for static languages, this algorithm
is tailored to the way dynamic languages work: the typing of code is
determined by the flow of data through it.  The algorithm works by
walking through a program, starting with the types of its input
values, and ``abstractly interpreting'' it: instead of applying the
code to values, it applies the code to types, following all branches
concurrently and tracking all possible states the program could be in,
including all the types each expression could assume.


The dataflow type inference algorithm allows programs to be
automatically annotated with type bounds without forcing the
programmer to explicitly specify types.  Yet, in dynamic languages it
is possible to write programs which inherently cannot be concretely
typed.  In such cases, dataflow type inference provides what bounds it
can, but these may be trivial and useless---i.e. they may not narrow
down the set of possible types for an expression at all.  However, the
design of Julia's programming model and standard library are such that
a majority of expressions in typical programs \textit{can} be
concretely typed.  Moreover, there is a positive correlation between
the ability to concretely type code and that code being
performance-critical.

%
%
%
%



A lesson of the numerical computing languages is that one must learn to
vectorize to get performance.  The mantra is ``for loops" are bad,
vectorization is good.  Indeed one can find the mantra on p.72 of the
``1998 Getting Started with Matlab manual'' (and other editions):

\begin{quotation}
Experienced Matlab  users like to say ``Life is too short to spend writing for loops."
\end{quotation}

It is not that ``for loops'' are inherently slow by themselves. The
slowness comes from the fact that in the case of most dynamic
languages, the system does not have access to the types of the
variables within a loop. Since programs often spend much of their time
doing repeated computations, the slowness of a particular operation
due to lack of type information is magnified inside a loop. This leads
to users often talking about ``slow for loops'' or ``loop overhead''.



In statically typed languages, full type information
is always available at compile time, allowing
compilation of a loop into a few machine instructions. This is
not the case in most dynamic languages, where the types are discovered
at run time, and the cost of determining the types and selecting the
right operation can run into hundreds or thousands of
instructions.

Julia has a transparent performance model. For example a
\verb+Vector{Float64}+ as in our example here, always has the same
in-memory representation as it would in C or Fortran; one can take a
pointer to the first array element and pass it to a C library function
using \verb+ccall+ and it will just work. The programmer knows exactly how
the data is represented and can reason about it. They know that a
\verb+Vector{Float64}+ does not require any additional heap
allocation besides the \verb+Float64+ values and that
arithmetic operations on these values will be machine arithmetic
operations. In the case of say, \verb+Complex128+, Julia stores
complex numbers in the same way as C or Fortran. Thus complex arrays
are actually arrays of complex values, where the real and imaginary
values are stored consecutively. Some systems have taken the path of
storing the real and imaginary parts separately, which leads to some
convenience for the user, at the cost of performance and
interoperability. With the \verb+immutable+ keyword, a programmer can
also define immutable data types, and enjoy the same benefits of
performance for composite types as for the more primitive number types
(bits types). This approach is being used to define many interesting
data structures such as small arrays of fixed sizes, which can have
much higher performance than the more general array data structure.

The transparency of the C data and performance models has been one of
the major reasons for C's long-lived success. One of the design goals
of Julia is to have similarly transparent data and performance
models. With a sophisticated type system and type inference, Julia
achieves both.

\section{Code selection: Run the right code at the right time}
\label{sec:select}

Code selection or code specialization from one point of view is the
opposite of code reuse enabled by abstraction.  Ironically, viewed
another way, it enables abstraction.  Julia allows users to overload
function names, and select code based on argument types.  This can
happen at the highest and lowest levels of the software stack.  Code
specialization lets us optimize for the details of the case at hand.
Code abstraction lets calling codes, probably those not yet even
written or perhaps not even imagined, work all the way through on
structures that may not have been envisioned by the original
programmer.

We see this as the ultimate realization of the famous 1908 quip that
\begin{quote}
Mathematics is the art of giving the same name to different things.
\end{quote}
by noted mathematician Henri Poincar\'{e}.\footnote{ A few versions of
  this quote are relevant to Julia's power of abstractions and
  numerical computing. They are worth pondering:
\begin{quote}
 It is the harmony of the different parts, their symmetry, and their
 happy adjustment; it is, in a word, all that introduces order, all
 that gives them unity, that enables us to obtain a clear
 comprehension of the whole as well as of the parts. Elegance may
 result from the feeling of surprise caused by the unlooked-for
 occurrence of objects not habitually associated. In this, again, it
 is fruitful, since it discloses thus relations that were until then
 unrecognized. {\bf Mathematics is the art of giving the same names to
   different things.}
\end{quote}
 http://www.nieuwarchief.nl/serie5/pdf/naw5-2012-13-3-154.pdf.
 and
 \begin{quote}
 One example has just shown us the importance of terms in mathematics;
 but I could quote many others. It is hardly possible to believe what
 economy of thought, as Mach used to say, can be effected by a
 well-chosen term. I think I have already said somewhere that {\bf
   mathematics is the art of giving the same name to different
   things}. It is enough that these things, though differing in
 matter, should be similar in form, to permit of their being, so to
 speak, run in the same mould. When language has been well chosen, one
 is astonished to find that all demonstrations made for a known object
 apply immediately to many new objects: nothing requires to be
 changed, not even the terms, since the names have become the same.
 \end{quote}
 {\tt
   http://www-history.mcs.st-andrews.ac.uk/Extras/Poincare\_Future.html
 }

}

In this upcoming section we provide examples of how plus can apply to
so many objects. Some examples are floating point numbers, or
integers.  It can also apply to sparse and dense matrices.  Another
example is the use of the same name, ``det", for determinant, for the
very different algorithms that apply to very different matrix
structures.  The use of overloading not only for single argument
functions, but for multiple argument functions is already a powerful
abstraction.

\subsection{Multiple Dispatch}
\label{sec:dispatch}

Multiple dispatch is the selection of a function implementation
based on the types of each argument of the function.
It is  not only a nice notation to remove a long list of ``case" statements,
but it is part of the reason for Julia's speed.
It is expressed in Julia by annotating the type of a function
argument in a function definition with the following syntax:  \verb+argument::Type+.

Mathematical notations that are often used in
 print are difficult to employ in programs.  For example, we can
 teach the computer some natural ways to multiply numbers and
 functions. Suppose that $a$ and $t$ are scalars, and $f$ and $g$ are
 functions, and we wish to define
\begin{enumerate}
\item   { \bf Number x Function \  = scale output:}  $a*g$ is the function that takes $x$ to $a*g(x)$
\vspace{-.05in}
\item {\bf Function x Number \ = scale argument :}  $f*t$ is the function that takes $x$ to $f(tx)$ and
\vspace{-.05in}
\item  {\bf Function x Function = composition of functions:} $f*g$ is the function that takes $x$ to $f(g(x))$.
\end{enumerate}

If you are a mathematician who does not program, you would not see the
fuss.  If you thought how you might implement this in your favorite
computer language, you might immediately see the benefit.  In Julia,
multiple dispatch makes all three uses of \verb+*+ easy to express:

\ja
*(a::Number,  g::Function)= x->a*g(x)   \>  \sh{    \# Scale output  }   \\
*(f::Function,t::Number)  = x->f(t*x)     \> \sh{  \# Scale  argument  } \\
*(f::Function,g::Function)= x->f(g(x))     \sh{  \# Function composition}
\end{jinput}

Here, multiplication is dispatched by the type of its first and second
 arguments.  It goes the usual way if both are numbers, but there are
 three new ways if one, the other, or both are functions.
\begin{figure}
  \centering
  \includegraphics[width=3in]{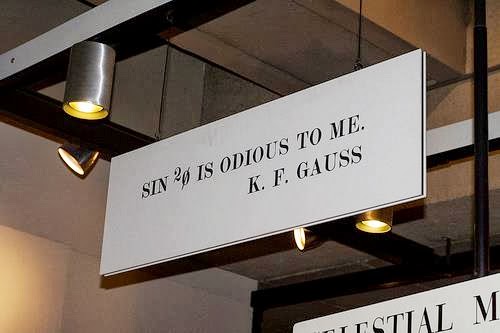}
  \caption{\label{fig:gauss}  Gauss quote hanging from the ceiling of the longstanding old Boston Museum of Science Mathematica Exhibit.
  }
\end{figure}

These definitions exist as part of a larger system of generic definitions,
which can be reused by later definitions.
Consider the case of the mathematician Gauss' preference for
$\sin^2 \phi $ to refer to $\sin(\sin(\phi))$ and not
$\sin(\phi)^2$ (writing ``$\sin^2(\phi)$ is odious to me, even
though Laplace made use of it."(Figure~\ref{fig:gauss}).)
By defining \verb+*(f::Function,g::Function)= x->f(g(x))+,
{\tt (f\^{}2)(x)} automatically computes $f(f(x))$ as Gauss
wanted. This is a consequence of a generic definition that evaluates
\verb+x^2+ as \verb+x*x+ no matter how \verb+x*x+ is defined.

This paradigm is a natural fit for numerical computing, since so
many important operations involve interactions among multiple
values or entities. Binary arithmetic operators are obvious examples,
but many other uses abound. The fact that the compiler can pick the
sharpest matching definition of a function based on its input types
helps achieve higher performance, by keeping the code execution paths
tight and minimal.

We have not seen this in the literature but it seems worthwhile to point out four possibilities:

\begin{enumerate}
\item Static single dispatch (not done)
\item Static multiple dispatch (frequent in static languages, e.g. C++ overloading)
\item Dynamic single dispatch  (Matlab's object oriented system might fall in this category though it has its own special characteristics)
\item Dynamic multiple dispatch (usually just called multiple dispatch).
\end{enumerate}

In Section \ref{sec:traditional} we discuss the comparison with
traditional object oriented approaches. Class-based object oriented
programming could reasonably be called dynamic single dispatch, and
overloading could reasonably be called static multiple dispatch.
Julia's dynamic multiple dispatch
approach is more flexible and adaptable while still retaining
powerful performance capabilities. Julia programmers often find that
dynamic multiple dispatch makes it easier to structure their programs
in ways that are closer to the underlying science.

\subsection{Code selection from bits to matrices}

Julia uses the same mechanism for code selection at all levels, from
the top to the bottom.


\begin{center}
\begin{tabular}{|c|c|c|} \hline
f &  Function &  Operand  Types \\\hline
Low Level ``+" &  Add Numbers&    \{Float , Int\}  \\
High Level ``+"  &  Add  Matrices &  \{Dense Matrix ,  Sparse Matrix\}  \\
 `` * " & Scale or Compose &  \{Function , Number \} \\ \hline
\end{tabular}
\end{center}

\subsubsection{Summing Numbers: Floats and Ints}
\label{sec:bits}

We begin at the lowest level.   Mathematically, integers are thought of as being special real numbers, but
on a computer,  an Int and a Float have two very different representations.
Ignoring for a moment that there are even many choices of Int and Float representations,
if we add two numbers,
 code selection based on numerical representation is taking place at a very low level.
 Most users are blissfully unaware of this code selection,  because it is hidden somewhere that is usually
off-limits to the user.
Nonetheless, one can follow the evolution of the high level code all the way down to the assembler level which ultimately would
reveal an ADD instruction for integer addition, and, for example,  the AVX\footnote{AVX: \color{red}A\color{black}danced \color{red}V\color{black}ector e\color{red}X\color{black}tension to the x86 instruction set} instruction  VADDSD\footnote{VADDSD: \color{red}{V}\color{black}ector \color{red}ADD S\color{black}calar \color{red}D\color{black}ouble-precision}  for floating point addition in the language
of x86 assembly level instructions.  The point being these are ultimately two different algorithms being called, one for a pair of Ints and one for a pair of Floats.

Figure \ref{fig:nativeadd} takes a close look at what a computer
 must do to perform  \verb-x+y- depending on whether (x,y) is (Int,Int), (Float,Float), or  (Int,Float) respectively.
In the first case, an integer add is called, while in the second case a float add is called.  In the last case,  a promotion of the int to float is called through the x86 instruction VCVTSI2SD\footnote{VCVTSI2SD: \color{red}{V}\color{black}ector
\color{red}{C}\color{black}{on}\color{red}{V}\color{black}{er}\color{red}{T}\color{black}{ Doubleword} (\color{red}S\color{black}calar) \color{red}{I}\color{black}{nteger to}\color{red}{(2) S}\color{black}{calar} \color{red}{D}\color{black}ouble Precision Floating-Point Value}, and then the float add follows.

It is instructive to build a  Julia simulator in Julia itself.
Let us define the aforementioned assembler instructions using Julia.

\ja
\sh {\# Simulate the assembly level add, vaddsd, and vcvtsi2sd commands}
\begin{verbatim}
add(x::Int       ,y::Int)     = x+y
vaddsd(x::Float64,y::Float64) = x+y
vcvtsi2sd(x::Int)             = float(x)
\end{verbatim}
\end{jinput}

\ja
\sh{\# Simulate Julia's definition of + using $\oplus$} \\
\sh{\# To type $\oplus$, type as in TeX,  \textbackslash oplus  and hit the  <tab> key} \\
$\oplus$\verb+(x::Int,    y::Int)     = add(x,y)+ \\
$\oplus$(x::Float64,y::Float64) = vaddsd(x,y) \\
$\oplus$\verb+(x::Int,    y::Float64) = vaddsd(vcvtsi2sd(x),y)+ \\
$\oplus$\verb+(x::Float64,y::Int)     = y+ $\oplus$  x\\
\end{jinput}

\ja
methods($\oplus$)
\jb
4 methods for generic function $\oplus$:\\
$\oplus$ (x::Int64,y::Int64) at In[23]:3 \\
$\oplus$ (x::Float64,y::Float64) at In[23]:4 \\
$\oplus$ (x::Int64,y::Float64) at In[23]:5\\
$\oplus$ (x::Float64,y::Int64) at In[23]:6
\jc

\begin{figure}
\caption{\label{fig:nativeadd} While assembly code may seem intimidating, Julia disassembles readily.  Armed with the {\tt code\_native}  command in Julia and perhaps a good list of
assembler commands such as  may be found on {\tt http://docs.oracle.com/cd/E36784\_01/pdf/E36859.pdf}
or
{\tt http://en.wikipedia.org/wiki/X86\_instruction\_listings}
 one can really learn to
see the details of
code selection in action at the lowest levels.
More importantly one can begin to understand that Julia is fast
because the assembly code produced is so tight.}

\hspace*{.14in}
\ja
f(a,b) = a + b
\jb
f (generic function with 1 method)
\jc

\hspace*{.14in}
\ja
\sh{\# Ints add with the x86  \underline{add}  instruction}
\vspace{-.06in}
\begin{verbatim}
@code_native f(2,3) \end{verbatim}
\jb
 push	RBP \\
 mov	RBP, RSP \\
 \sh{add}	RDI, RSI \\
 mov	RAX, RDI \\
 pop	RBP \\
 ret
\jc

\hspace*{.14in}
\ja
\sh{\# Floats add, for example,  with the x86  \underline{vaddsd}  instruction}
\vspace{-.06in}
\begin{verbatim}
@code_native f(1.0,3.0)
\end{verbatim}
\jb
 push	RBP \\
 mov	RBP, RSP\\
 \sh{vaddsd} XMM0, XMM0, XMM1\\
 pop	RBP\\
 ret
 \jc

\hspace*{.14in}
\ja
\sh{\# Int + Float requires a convert to scalar double precision, hence \\  \#  the  x86 \underline{vcvtsi2sd} instruction}
\vspace{-.06in}
\begin{verbatim}
@code_native f(1.0,3)
\end{verbatim}
\jb

 push	RBP  \\
 mov	RBP, RSP  \\
 \sh{vcvtsi2sd}	XMM1, XMM0, RDI \\
 \sh{vaddsd}	XMM0, XMM1, XMM0 \\
 pop	RBP \\
 ret

 \jc

\end{figure}

\subsubsection{Summing Matrices: Dense and Sparse}
\label{sec:summing}
We now move to a much higher level: matrix addition.
The versatile ``+" symbol lets us add matrices.
Mathematically, sparse matrices are thought of as being special matrices
with enough zero entries.
On a computer, dense matrices are (usually)  contiguous blocks of data with a few parameters attached,
while sparse matrices (which may be stored in many ways) require storage of index information one way or another.
If we add two matrices, code selection must take place depending on whether the summands are (dense,dense),
(dense,sparse), (sparse,dense) or (sparse,sparse).

While this is at a much higher level, the basic pattern is unmistakably the same as that
of Section \ref{sec:bits}.
We show how to use a dense algorithm in the implementation
of $\oplus$ when either $A$ or $B$ (or both) are dense.  A sparse algorithm
is used when both $A$ and $B$ are sparse.

\ja
\sh{\# Dense + Dense} \\
$\oplus$(A::Matrix,              B::Matrix)               =\\
\hspace*{0.3in}  [A[i,j]+B[i,j] for i in 1:size(A,1),j in 1:size(A,2)] \\
\sh{\# Dense + Sparse} \\
$\oplus$(A::Matrix,              B::AbstractSparseMatrix) = A $\oplus$ full(B) \\
\sh{\#  Sparse + Dense} \\
$\oplus$(A::AbstractSparseMatrix,B::Matrix)          \     = B $\oplus$ A \sh{\# Use Dense + Sparse}  \\
\sh{\# Sparse + Sparse is best written using the long form function definition:}
function $\oplus$(A::AbstractSparseMatrix, B::AbstractSparseMatrix)
\vspace{-0.06in}
\begin{verbatim}
    C=copy(A)
    (i,j)=findn(B)
    for k=1:length(i)
        C[i[k],j[k]]+=B[i[k],j[k]]
    end
    return C
end
\end{verbatim}
\end{jinput}

We now have eight methods for the function $\oplus$, four for the low level sum,
and four more for the high level sum.

\ja
methods($\oplus$)
\jb
8 methods for generic function $\oplus$:\\
$\oplus$ (x::Int64,y::Int64) at In[23]:3 \\
$\oplus$ (x::Float64,y::Float64) at In[23]:4 \\
$\oplus$ (x::Int64,y::Float64) at In[23]:5\\
$\oplus$ (x::Float64,y::Int64) at In[23]:6 \\
$\oplus$  (A::Array\{T,2\},B::Array\{T,2\}) at In[29]:1 \\
$\oplus$ (A::Array\{T,2\},B::AbstractSparseArray\{Tv,Ti,2\}) at In[29]:1 \\
$\oplus$ (A::AbstractSparseArray\{Tv,Ti,2\},B::Array\{T,2\}) at In[29]:1 \\
$\oplus$  (A::AbstractSparseArray\{Tv,Ti,2\},B::AbstractSparseArray\{Tv,Ti,2\}) at In[29]:2
\jc

\subsection{The many levels of code selection}

In Julia as in mathematics, functions are as important as the data
they operate on, their arguments. Perhaps even more so.
 We can create a new
function \verb+foo+ and gave it six definitions depending on the
combination of types. In the following example we
sensitize unfamiliar readers with terms  from
computer science language research.
It is not critical that these terms be understood all at once.

\begin{jinput}
\sh {\# Define a} generic function \sh{with 6 methods. Each method is itself a} \\
\sh{ \# function. In Julia generic functions are far more convenient than the } \\
\sh{ \# multitude of  case statements seen in other languages.
When Julia sees} \\
\sh{\#}  foo, \sh{  it decides which method to use, rather than first seeing and deciding} \\
\sh{\#  based on the type.}
\begin{verbatim}
foo() = "Empty input"
foo(x::Int) = x
foo(S::String) = length(S)
foo(x::Int, S::String) = "An Int and a String"
foo(x::Float64,y::Float64) = sqrt(x^2+y^2)
foo(a::Any,b::String)= "Something more general than an Int and a String"
\end{verbatim}
\sh{\# The function name} foo  \sh{is overloaded. This is an example of} polymorphism. \\
\sh{\# In the jargon of computer languages this is called} ad-hoc polymorphism. \\
\sh{\# The} multiple dynamic dispatch \sh{idea captures the notion that the generic } \\
\sh{\# function is deciphered dynamically at runtime.  One of the six choices} \\
\sh{\# will be made or an error will occur.}
\end{jinput}
\begin{joutput}
foo (generic function with 6 methods)
\end{joutput}

Any one instance of \verb+foo+ is known as a method or function.  The
collection of six methods is referred to as a {\bf generic function}.
The word ``polymorphism" refers to the use of the same name
(foo, in this example) for functions with different types.
Contemplating the Poincar\'{e} quote in Footnote 5, it is handy to
reason about everything that you are giving the same name.  In real
life coding, one tends to use the same name when the abstraction makes
a great deal of sense. That we use "+" for ints,floats, dense
matrices, and sparse matrices is the same name for different things.
Methods are grouped into generic functions.

While mathematics is the art of giving the same name to seemingly
different things, a computer has to eventually execute the right
program in the right circumstance. Julia's code selection operates at
multiple levels in order to translate a user's abstract ideas into
efficient execution.  A generic function can operate on several
arguments, and the method with the most specific signature matching
the arguments is invoked. It is worth crystallizing some key aspects
of this process:

\begin{enumerate}
\item The same name can be used for different functions in different
  circumstances. For example, \verb+select+ may refer to the selection
  algorithm for finding the $k^{th}$ smallest element in a list, or to
  select records in a database query, or simply as a user-defined
  function in a user's own program. Julia's namespaces allow the
  usage of the same vocabulary in different circumstances in a simple
  way that makes programs easy to read.
\item A collection of functions that represent the same idea but
  operate on different structures are naturally referred to by the
  same name. Which method is called is based entirely on the types of
  all the arguments - this is multiple dispatch. The function
  \verb+det+ may be defined for all matrices at an abstract
  level. However, for reasons of efficiency, Julia defines different
  methods for different types of matrices, depending on whether they
  are dense or sparse, or if they have a special structure such as
  diagonal or tridiagonal.
\item Within functions that operate on the same structure, there may
  be further differences based on the different types of data contained
  within. For example, whether the input is a vector of Float64 values
  or Int32 values, the norm is computed in the same exact way,
  with a common body of code, but the compiler is able to generate
  different executable code from the abstract specification.
\item Julia uses the same mechanism of code selection at the lowest
  and highest levels - whether it is performing operations on matrices or
  operations on bits. As a result, Julia is able to optimize the whole
  program, picking the right method at the right time, either at
  compile-time or run-time.
\end{enumerate}

\subsection{Is ``code selection"  just traditional object oriented programming?}
\label{sec:traditional}

\begin{figure}

\centering
\hspace*{1in}\begin{minipage}{5in}
\begin{shaded}
\sh{\textbackslash * Polymorphic Java Example. Method defined by types of two arguments.  *\textbackslash} \\
\begin{verbatim}
public class OverloadedAddable {

	   public int    addthem(int i, int f} {
	      return i+f;
	   }

	   public double addthem(int i, double f} {
	      return i+f;
	   }

	   public double addthem(double i, int f} {
	      return i+f;
	   }

	   public double addthem(double i,  double  f} {
	      return i+f;
	   }

}
\end{verbatim}}
\end{shaded}
\end{minipage} \caption{\label{fig:java} Advantages of Julia: It is
  true that the above Java code is polymorphic based on the types of
  the two arguments.
  (``Polymorphism" is the use of the same name for a function
  that may have different type arguments.)
   However, in Java if the method {\tt addthem} is
  called, the types of the arguments must be known at compile time.
  This is static dispatch. Java is also encumbered by encapsulation:
  in this case {\tt addthem} is encapsulated inside the {\tt
    OverloadedAddable} class.  While this is considered a safety
  feature in Java culture, it becomes a burden for numerical
  computing.  }
\end{figure}

The method to be executed in Julia is not chosen by only one argument,
which is what happens in the case of single dispatch, but through
multiple dispatch that considers the types of all the arguments.
Julia is not encumbered by the encapsulation restrictions (class based
methods) of most object oriented languages. The generic functions play
a more important role than the data types. Some call this ``verb"
based languages as opposed to most object oriented languages being
``noun" based.  In numerical computing, it is the concept of ``solve
$Ax=b$" that often feels more primary, at the highest level, rather
than whether the matrix $A$ is full, sparse, or structured.  Readers
familiar with Java might think, "So what? One can easily create
methods based on the types of the arguments".  An example is provided
in Figure \ref{fig:java}. However a moment's thought shows that the
following dynamic situation in Julia is impossible to express in Java:

\ja
\sh {\# It is possible for a static compiler to know that x,y are Float} \\
x = rand(Bool) ?  1.0 : 2.0 \\
y = rand(Bool) ?  1.0 : 2.0 \\
x+y \\

\sh {\# It is impossible to know until runtime if x,y are Int or Float} \\
x = rand(Bool) ? 1 : 1.0 \\
y = rand(Bool) ? 1 : 1.0 \\
x+y
\end{jinput}

\vspace{0.1in}

Readers are familiar with single dispatch mechanism, as in Matlab.  It
is unusual in that it is not completely class based, as the code
selection is based on Matlab's own custom hierarchy.  In Matlab the
leftmost object has precedence, but user-defined classes have
precedence over built-in classes.  Matlab also has a mechanism to
create a custom hierarchy.

Julia generally shuns the notion of ``built-in" vs.\ ``user-defined"
preferring to focus on the method to be performed based on the
combination of types, and obtaining high performance as a byproduct.
A high level library writer, which we do not distinguish from any
user, has to match the best algorithm for the best input structure.  A
sparse matrix would match to a sparse routine, a dense matrix to a
dense routine.  A low level language designer has to make sure that
integers are added with an integer adder, and floating points are
added with a float adder.  Despite the very different levels, the
reader might recognize that deep down, these are both examples of code
being selected to match the structure of the problem.

Readers familiar with object-oriented paradigms such as C++ or Java
are most likely familiar with the approach of encapsulating methods
inside classes.
Julia's more general multiple dispatch mechanism (also known as
generic functions, or multi-methods) is a paradigm where methods are
defined on combinations of data types (classes)
Julia has proven that this is remarkably well suited for numerical
computing.

A class based language might express the sum of a sparse matrix with a
full matrix as follows: {\tt A\_sparse\_matrix.plus(A\_full\_matrix)}.
Similarly it might express indexing as \newline {\tt
  A\_sparse\_matrix.sub(A\_full\_matrix)} .  If a tridiagonal were
added to the system, one has to find the method {\tt plus} or {\tt
  sub} which is encapsulated in the sparse matrix class, modify it and
test it. Similarly, one has to modify every full matrix method, etc.
We believe that class-based methods, which can be taken quite far, are
not sufficiently powerful to express the full gamut of abstractions in
scientific computing.  Further, the burdens of encapsulation create a
wall around objects and methods that are counterproductive for
numerical computing.  \newline \hspace*{.2in} The generic function
idea captures the notion that a method for a general operation on
pairs of matrices may exist (e.g. ``+'') but if a more specific
operation is possible (e.g. ``+'' on sparse matrices, or ``+'' on a
special matrix structure like Bidiagonal), then the more specific
operation is used.  We also mention indexing as another example, Why
should the indexee take precedence over the index?

%

\subsection{Quantifying the use of multiple dispatch}

In~\cite{juliaarray} we performed an analysis to substantiate the
claim that multiple dispatch, an esoteric idea for numerical computing from computer
languages, finds its killer application in scientific computing.
We wanted to answer for ourselves the question of whether there
was really anything different about how \julia\ uses multiple
dispatch.

Table \ref{dispatchratios} gives an answer in terms of Dispatch ratio (DR),
Choice ratio (CR). and Degree of specialization (DoS).
While multiple dispatch is an idea that has been circulating for some time,
its application to numerical computing appears to have significantly favorable
characteristics compared to previous applications.

To quantify how heavily a language feature is used,
we use the following metrics for evaluating the extent of multiple
dispatch \cite{Muschevici:2008}:

\begin{enumerate}
\item Dispatch ratio (DR): The average number of methods in a generic
 function.
\item Choice ratio (CR): For each method, the total number of methods
 over all generic functions it belongs to, averaged over all
 methods. This is essentially the sum of the squares of the number of
 methods in each generic function, divided by the total number of
 methods. The intent of this statistic is to give more weight to
 functions with a large number of methods.
\item Degree of specialization (DoS): The average number of
 type-specialized arguments per method.
\end{enumerate}

Table~\ref{dispatchratios} shows the mean of each metric over the
entire Julia \code{Base} library, showing a high degree of multiple
dispatch compared with corpora in other languages
\cite{Muschevici:2008}.  Compared to most multiple dispatch systems,
\julia\ functions tend to have a large number of definitions. To see
why this might be, it helps to compare results from a biased sample of
common operators. These functions are the most obvious candidates for
multiple dispatch, and as a result their statistics climb
dramatically. \julia\ is focused on numerical computing, and so is
likely to have a large proportion of functions with this character.

\begin{table}
\label{dispatchratios}
\begin{center}
\begin{tabular}{|l|r|r|r|}\hline
Language & DR & CR & DoS \\
\hline \hline
Gwydion    & 1.74 & 18.27 & 2.14 \\
\hline
OpenDylan  & 2.51 & 43.84 & 1.23 \\
\hline
CMUCL      & 2.03 &  6.34 & 1.17 \\
\hline
SBCL       & 2.37 & 26.57 & 1.11 \\
\hline
McCLIM     & 2.32 & 15.43 & 1.17 \\
\hline
Vortex     & 2.33 & 63.30 & 1.06 \\
\hline
Whirlwind  & 2.07 & 31.65 & 0.71 \\
\hline
NiceC      & 1.36 &  3.46 & 0.33 \\
\hline
LocStack   & 1.50 &  8.92 & 1.02 \\
\hline
\julia\      & 5.86 & 51.44 & 1.54 \\
\hline
\julia\ operators & 28.13 & 78.06 & 2.01 \\
\hline
\end{tabular}
\end{center}
\caption{
A comparison of \julia\ (1208 functions exported from the \code{Base} library)
to other languages with multiple dispatch.
The ``\julia\ operators'' row describes 47 functions with special syntax
(binary operators, indexing, and concatenation).
Data for other systems are from \cite{Muschevici:2008}.
The results indicate that \julia\ is using multiple dispatch far more heavily
than previous systems.  }
\end{table}

\subsection{Case Study for Numerical Computing}

The complexity of linear algebra software has been nicely captured in the
context of LAPACK and ScaLAPACK by Demmel and Dongarra, et.al.,
\cite{lawn181} and reproduced verbatim here:

\begin{verbatim}
(1) for all linear algebra problems
     (linear systems, eigenproblems, ...)
(2)     for all matrix types
            (general, symmetric, banded, ...)
(3)           for all data types
                  (real, complex, single, double, higher precision)
(4)                  for all machine architectures
                       and communication topologies
(5)                        for all programming interfaces
(6)                             provide the best algorithm(s) available in terms of
                                  performance and accuracy (``algorithms" is plural
                                  because sometimes no single one is always best)
\end{verbatim}


In the language of Computer Science, code reuse is about taking
advantage of polymorphism.  In the general language of mathematics
it's about taking advantage of abstraction, or the sameness of two
things.  Either way, programs are efficient, powerful, and
maintainable if programmers are given powerful mechanisms to reuse
code.


Increasingly, the applicability of linear algebra has gone well beyond
the LAPACK world of floating point numbers.  These days linear algebra
is being performed on, say, high precision numbers, integers, elements
of finite fields, or rational numbers. There will always be a special
place for the BLAS, and the performance it provides for floating point
numbers.  Nonetheless, linear algebra operations transcend any one
data type.  One must be able to write a general implemenation and as
long as the necessary operations are available, the code should just
work \cite{andreas}.  That is the power of code reuse.

\subsubsection{Determinant: Simple Single Dispatch}

In traditional numerical computing there were people with special skills known as
library writers.  Most users were, well, just users of libraries.  In this case study,
we show how anybody can dispatch a new determinant function based solely
on the type of the argument.

For triangular and diagonal structures the obvious formulas are used.
  For general matrices, the programmer will compute a QR decomposition
 of the matrix and find the determinant as the product of the diagonal
 elements of $R$.\footnote{LU is more efficient. We simply wanted to
 illustrate other ways are possible.}  For symmetric tridiagonals the
 usual 3-term recurrence formula\cite{threetermrecurrence} is used.  (The first four are defined
 as one line functions; the symmetric tridiagonal uses the long form.)

\begin{jinput}
\sh{\# Simple determinants defined using the short form for functions} \\
newdet(x::Number)     = x \\
newdet(A::Diagonal )  = prod(diag(A))   \\
newdet(A::Triangular) = prod(diag(A))   \\
newdet(A::Matrix) = -prod(diag(qrfact(full(A))[:R]))*(-1)\^{}size(A,1)  \\

\sh{\# Tridiagonal determinant defined using the long form for functions} \\
function newdet(A::SymTridiagonal)  \\
\, \sh{\# Assign c and d as a pair }\\
    \,  c,d = 1, A[1,1]   \\
    \,  for i=2:size(A,1)    \\
\,  \,     {\sh{ \#  temp=d, d=the expression, c=temp}} \\
     \, \,    c,d = d, d*A[i,i]-c*A[i,i-1]{}\^ {}{2}      \\
       \,   end \\
   \,  d \\
end  \\
\end{jinput}

We have illustrated a mechanism to select a determinant formula at runtime based on the type of
the input argument.  If Julia knows an argument type early, it can make use of this information
for performance.  If it does not, code selection can still happen, at runtime.  The reason why Julia
can still perform well is that once code selection based on type  occurs, Julia can return to performing well
once inside the method.

\subsubsection{A Symmetric Arrow Matrix Type}
\label{sec:arrow}
In the field of Matrix Computations, there are matrix structures and operations on these matrices.  In Julia, these
structures exist as Julia types.
Julia has a number of predefined matrix structure types: (dense) \verb+Matrix+, (compressed sparse column) \verb+SparseMatrixCSC+,
\verb+Symmetric+, \verb+Hermitian+, \verb+SymTridiagonal+,
\verb+Bidiagonal+, \verb+Tridiagonal+,  \verb+Diagonal+,
and \verb+Triangular+ are all examples of Julia's matrix structures.

The operations on these matrices exist as Julia functions.  Familiar examples of operations are indexing, determinant, size,
and matrix addition.  Since matrix addition takes two arguments, it may be necessary to reconcile two different types when
computing the sum.

Some languages do not allow you to extend their built in functions and types.
This ability is known
as external dispatch.   In the following example, we illustrate how the user can add symmetric
arrow matrices to the system, and then add a specialized \verb+det+ method to compute
the determinant of a symmetric arrow matrix efficiently.
We build on the symmetric arrow type introduced in Section \ref{sec:firstclass}.

\ja
\sh{\# Define a Symmetric Arrow Matrix Type} \\
immutable SymArrow\{T\}  <: AbstractMatrix\{T\} \\
 \,          dv::Vector\{T\}                     \sh{   \# diagonal  }\\
 \,          ev::Vector\{T\}                     \sh{   \# 1st row{}[2:n{}] }\\
 \>      end
\end{jinput}

\ja
\sh {\# Define its size} \\
importall Base \\
size(A::SymArrow, dim::Integer) = size(A.dv,1) \\
size(A::SymArrow)= size(A,1), size(A,1)
\jb
size (generic function with 52 methods)
\jc

\ja
\sh {\# Index into a SymArrow} \\
 function getindex(A::SymArrow,i::Integer,j::Integer) \\
\,           if     i==j; return A.dv[i]   \\
\, \>         elseif i==1; return A.ev{}[j-1{}]  \\
\,  \>       elseif j==1; return A.ev{}[i-1{}] \\
\,    \>         else         return zero(typeof(A.dv{}[1{}]))  \\
\,           end \\
    end
       \jb
getindex (generic function with 168 methods)
\jc

\ja
\sh {\# Dense version of SymArrow} \\
full(A::SymArrow) =[A[i,j] for i=1:size(A,1), j=1:size(A,2)]
\jb
full (generic function with 17 methods)
\jc

\ja
\sh {\# An example} \\
S=SymArrow({}[1,2,3,4,5{}],{}[6,7,8,9{}])
\jb
5x5 SymArrow\{Int64\}: \\
\sh{ 1  6  7  8  9} \\
\sh{ 6  2}  0  0  0 \\
 \sh{7}  0  \sh{3}  0  0 \\
 \sh{8}  0  0  \sh{4}  0 \\
 \sh{9}  0  0  0  \sh{5}
 \jc

\ja
\sh{\# det for SymArrow  (external dispatch example)} \\
     function exc\_prod(v)  \sh{\# prod(v)/v[i] } \\
\,        [prod(v{}[{}[1:(i-1),(i+1):end{}]{}]) for i=1:size(v,1)]  \\
       end \\
       \sh{\# det for SymArrow  formula} \\
      det(A::SymArrow) = prod(A.dv)-sum(A.ev.\textasciicircum 2.*exc\_prod(A.dv{}[2:end{}]))
       \jb
       det (generic function with 17 methods)
\jc

The above julia code uses the  special formula
$$\det(A)=\prod_{i=1}^n d_i - \sum_{i=2}^n  e_i^2 \prod_{2 \le j \ne i \le n} d_j ,$$
valid for symmetric arrow matrices with diagonal $d$ and first row starting with the second entry $e$.

 In some numerical computing languages,
a  function might begin with a lot of argument checking to pick which algorithm to use.  In Julia, one creates
a number of {\it methods.}  Thus \verb+newdet+ on a diagonal is one method for \verb+newdet+, and \verb+newdet+
on a triangular matrix is a second method.  \verb+det+ on a \verb+SymArrow+ is a new method for \verb+det+. (See Section 4.6.1.)
Code is selected, in advance if the compiler knows the type, otherwise the code is selected at run time.  The selection
of code is known as {\it dispatch.}

We have seen a number of examples of code selection for single dispatch, i.e.,
the selection of code based on the type of a single argument.
We can now turn to a powerful feature,  Julia's multiple dispatch mechanism.
Now that we have created a symmetric arrow matrix, we might want to add it
to all possible matrices of all types.  However, we might notice that
a symmetric arrow plus a diagonal does not require operations on full dense matrices.

The code below starts with the most general case, and then allows for specialization
for the symmetric arrow and diagonal sum:

\ja
\sh{\# SymArrow + Any Matrix: (Fallback: add full dense arrays )} \\
+(A::SymArrow, B::Matrix) = full(A)+B  \\
+(B::Matrix, A::SymArrow) = A+B  \\
\sh{\# SymArrow + Diagonal:  (Special case: add diagonals, copy off-diagonal) }\\
+(A::SymArrow, B::Diagonal) = SymArrow(A.dv+B.diag,A.ev) \\
+(B::Diagonal, A::SymArrow) = A+B
\end{jinput}

\begin{figure}
  \centering
  \includegraphics[width=6.5in]{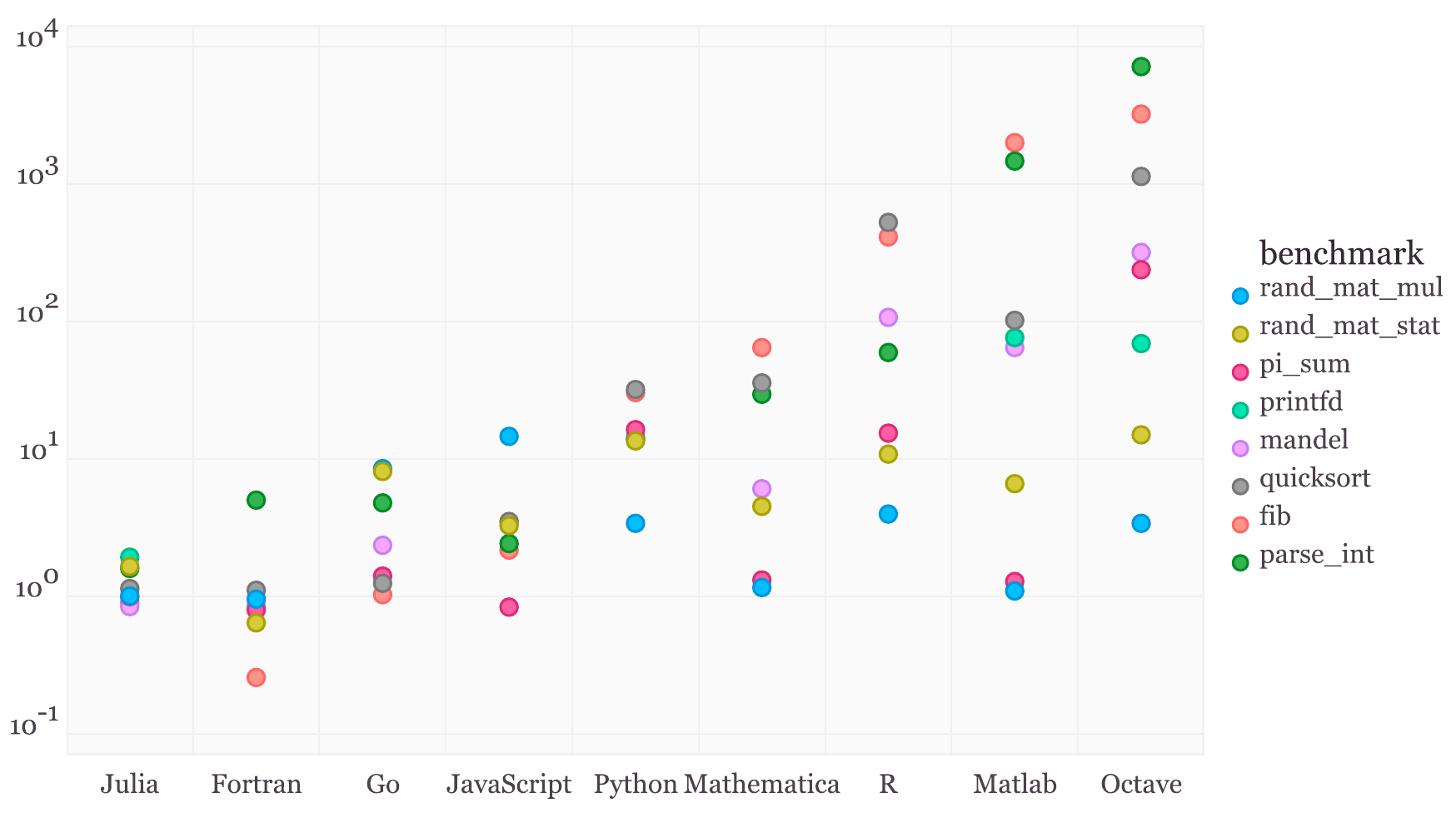}
  \caption{\label{fig:performance}{Performance comparison of various language performing simple micro-benchmarks. Benchmark execution time relative to C.  (Smaller is better, C performance = 1.0).}}
\end{figure}

\section{Leveraging language design for high performance libraries}
\label{sec:lang}

Seemingly innocuous design choices in a language can have profound,
pervasive performance implications.  These are often overlooked in
languages that were not designed from the beginning to be able to
deliver excellent performance.  Other aspects of language and library
design affect the usability, composability, and power of the provided
functionality.

\subsection{Integer arithmetic}

A simple but crucial example of a performance-critical language design
choice is integer arithmetic.  Julia uses machine arithmetic for
integer computations.

Consider what happens if we make the number of loop iterations fixed:

\ja
\sh{\# 10 Iterations of f(k)=5k-1 on integers} \\
 function g(k) \\
\,         for i = 1:10  \\
\, \>           k = f(k) \\
\,         end \\
\,         k  \\
       end
       \jb
g (generic function with 2 methods)
\jc

\ja
code\_native(g,(Int,))
\jb
Source line: 3 \\
 \>   push    RBP \\
\>    mov RBP, RSP  \\
Source line: 3 \\
\>    \sh{imul}    RAX, RDI, 9765625 \\
  \>   \sh{add} RAX, -2441406 \\
Source line: 5 \\
\>    pop RBP \\
 \>   ret \\
\jc

Because the compiler knows that integer addition and multiplication are associative and that multiplication distributes over addition it can optimize the entire loop down to just a multiply and an add.  Indeed, if $f(k)=5k-1$, it is true that the tenfold iterate $f^{(10)}(k)=-2441406 + 9765625 k$.

\subsection{A powerful approach to linear algebra}

We describe how the Julia language features have been used to provide
a powerful approach to linear algebra\cite{andreas}.

\subsubsection{Matrix factorizations}

  For decades, orthogonal matrices have been represented internally as
  products of Householder matrices stored in terms of vectors, and displayed for humans as matrix
  elements.  $LU$ factorizations are often performed in place,
  storing the $L$ and $U$ information together in the data locations
  originally occupied by $A$.  All this speaks to the fact that matrix
  factorizations deserve to be first class objects in a linear algebra
  system.

  In Julia, thanks to the contributions of Andreas Noack Jensen \cite{andreas} and
  many others, these structures are indeed first class objects.  The
  structure \verb+QRCompactWY+ holds a compact $Q$ and an $R$ in memory.
  Similarly an \verb+LU+ holds an $L$ and $U$ in packed form in memory.
  Through the magic of multiple dispatch, we can solve linear systems,
  extract the pieces, and do least squares directly on these
  structures.

  The $QR$ example is even more fascinating.  Suppose one computes $QR$ of
  a $4 \times 3$ matrix.  What is the size of $Q$?  The right answer, of
  course, is that it depends: it could be $4 \times 4$ or $4 \times 3$.  The
  underlying representation is the same.

  In Julia one can compute \verb+Aqr = qrfact(rand(4,3))+.  Then one
  extract Q  from the factorization  with  \verb+Q=Aqr[:Q]+.  This $Q$ retains its clever underlying structure
  and therefore is efficient and applicable when multiplying vectors
  of length 4 or length 3, contrary to the rules of freshman linear
  algebra, but welcome in numerical libraries for saving space and
  faster computations.

\ja

A=[1 2 3 \\
1 2 1 \\
      1 0 1  \\
     1 0 -1] \\

 Aqr = qrfact(A)); \\
 Q = Aqr[:Q] \jb
\begin{verbatim}
4x4 QRCompactWYQ{Float64}:
 -0.5  -0.5  -0.5
 -0.5  -0.5   0.5
 -0.5   0.5  -0.5
 -0.5   0.5   0.5
 \end{verbatim}
 \jc

\ja Q*[1,0,0,0] \jb
\begin{verbatim}
4-element Array{Float64,1}:
 -0.5
 -0.5
 -0.5
 -0.5
\end{verbatim}
 \jc

\ja Q*[1, 0, 0] \jb \begin{verbatim}
4-element Array{Float64,1}:
 -0.5
 -0.5
 -0.5
 -0.5
\end{verbatim}
\jc

\subsubsection{User-extensible wrappers for BLAS and LAPACK}

The tradition in linear algebra is to leave the coding to LAPACK
writers, and call LAPACK for speed and accuracy.  This has worked
fairly well, but Julia exposes considerable opportunities for
improvement.

Firstly, all of LAPACK is available to Julia users, not just the most
common functions. All LAPACK wrappers are implemented fully in Julia
code, using
\verb+ccall+\footnote{\url{http://docs.julialang.org/en/latest/manual/calling-c-and-fortran-code/}},
which does not require a C compiler, and can be called directly from
the interactive Julia prompt.  This makes it easy for users to
contribute LAPACK functionality, and that is how Julia's LAPACK
functionality has grown bit by bit. Wrappers for missing LAPACK
functionality can also be added by users in their own code.

Consider the following example that implements the Cholesky
factorization by calling LAPACK's \code{xPOTRF}. It uses Julia's
metaprogramming facilities to generate four functions, each
corresponding to the \code{xPOTRF} functions for \code{Float32}, \code{Float64}, \code{Complex64},
and \code{Complex128} types. The actual call to the Fortran functions is
wrapped in {\tt ccall}. Finally, the {\tt chol} function provides a
user-accessible way to compute the factorization.  It is easy to modify
the template below for any LAPACK call.

\begin{jinput}
\sh {\# Generate  calls to LAPACK's Cholesky for double, single, etc.}  \\
\sh {\# xPOTRF refers to POsitive definite TRiangular Factor}  \\
\sh{\# LAPACK signature: SUBROUTINE DPOTRF( UPLO, N, A, LDA, INFO ) }  \\
\sh{\# LAPACK documentation:}
\vspace{-0.05in}
\begin{verbatim}
*  UPLO    (input) CHARACTER*1
*          = 'U':  Upper triangle of A is stored;
*          = 'L':  Lower triangle of A is stored.
*  N       (input) INTEGER
*          The order of the matrix A.  N >= 0.
*  A       (input/output) DOUBLE PRECISION array, dimension (LDA,N)
*          On entry, the symmetric matrix A.  If UPLO = 'U', the leading
*          N-by-N upper triangular part of A contains the upper
*          triangular part of the matrix A, and the strictly lower
*          triangular part of A is not referenced.  If UPLO = 'L', the
*          leading N-by-N lower triangular part of A contains the lower
*          triangular part of the matrix A, and the strictly upper
*          triangular part of A is not referenced.
*          On exit, if INFO = 0, the factor U or L from the Cholesky
*          factorization A = U**T*U or A = L*L**T.
*  LDA     (input) INTEGER
*          The leading dimension of the array A.  LDA >= max(1,N).
*  INFO    (output) INTEGER
*          = 0:  successful exit
*          < 0:  if INFO = -i, the i-th argument had an illegal value
*          > 0:  if INFO = i, the leading minor of order i is not
*                positive definite, and the factorization could not be
*                completed.
\end{verbatim} \\
\sh {\# Generate Julia method potrf!}  \\
\vspace{-.1in}
\verb+ for  (potrf,  elty) in + \sh{\# Run through 4 element types}
\vspace{.05in}
 \begin{verbatim}
     ((:dpotrf_,:Float64),
      (:spotrf_,:Float32),
      (:zpotrf_,:Complex128),
      (:cpotrf_,:Complex64))
      \end{verbatim}
      \vspace{-.2in}
      \sh {\# Begin function potrf!}
\vspace{-.07in}
           \begin{verbatim}
     @eval begin
         function potrf!(uplo::Char, A::StridedMatrix{$elty})
             lda = max(1,stride(A,2))
             lda==0 && return A, 0
             info = Array(Int, 1)
             \end{verbatim}
                   \vspace{-.2in}
             \sh {\# Call to LAPACK:ccall(LAPACKroutine,Void,PointerTypes,JuliaVariables)}
             \vspace{-.07in}
             \begin{verbatim}
             ccall(($(string(potrf)),:liblapack), Void,
                    (Ptr{Char}, Ptr{Int}, Ptr{$elty}, Ptr{Int}, Ptr{Int}),
\end{verbatim}
\vspace{-.07in}
\verb+                    +\sh{\ \ \ \&uplo, \ \ \ \&size(A,1), \ \ \  A, \ \ \ \ \ \  \ \&lda, \ \  info})
\vspace{-.07in}
                    \begin{verbatim}
             return A, info[1]
          end
     end
end

chol(A::Matrix) = potrf!('U', copy(A))
\end{verbatim}

\end{jinput}

\subsection{High Performance Polynomials and Special Functions with Macros}
\label{subsection:macros}

Julia has a macro system that provides easy custom code generation,
bringing a level of performance that is otherwise difficult to
achieve.
A macro is a function that runs at parse-time, and takes parsed symbolic expressions in and returns transformed symbolic expressions out, which are inserted into the code for later compilation.

For example, a library developer implemented an
\verb+@evalpoly+ macro that uses Horner's rule to evaluate polynomials
efficiently.
Consider

\ja
@evalpoly(10,3,4,5,6)
\end{jinput}
\noindent which returns 6543 (the polynomial $3+4x+5x^2+6x^3$, evaluated at $10$ with Horner's rule).
Julia allows us to see the inline generated code with the command

\ja
macroexpand(:@evalpoly(10,3,4,5,6))
\end{jinput}

We reproduce the key lines below
\begin{joutput}
\#471\#t = 10  \sh{\# Store 10 into a variable named \#471\#t } \\
Base.Math.+(3,Base.Math.*(\#471\#t,Base.Math.+(4,Base.Math.*
            (\#471\#t,Base.Math.+(5,Base.Math.*(\#471\#t,6))))
           ))
\jc

This code-generating macro only needs to produce the correct symbolic
structure, and Julia's compiler handles the remaining details of fast
native code generation. Since polynomial evaluation is so important
for numerical library software it is critical that users can evaluate
polynomials as fast as possible.
The overhead of implementing an explicit loop, accessing coefficients in an array, and possibly a subroutine call (if it is not inlined), is substantial compared to just inlining the whole polynomial evaluation.

Steven Johnson reports in his EuroSciPy notebook\footnote{\url{https://github.com/stevengj/Julia-EuroSciPy14/blob/master/Metaprogramming.ipynb}}
\begin{quotation}
This is precisely how erfinv is implemented in Julia (in single and double precision), and is 3
to 4 times  faster than the compiled (Fortran?) code in Matlab, and 2 to 3 times  faster than the compiled (Fortran Cephes) code used in SciPy.

The difference (at least in Cephes) seems to be mainly that they have explicit arrays of polynomial coefficients and call a subroutine for Horner's rule, versus inlining it via a macro.
\end{quotation}

Johnson also used the same trick in his implementation of the
digamma special function for complex arguments\footnote{\url{https://github.com/JuliaLang/julia/issues/7033}} following an idea of Knuth:
\begin{quote}
  As described in Knuth TAOCP vol.\ 2, sec.\ 4.6.4, there is actually an
  algorithm even better than Horner's rule for evaluating polynomials
  p(z) at complex arguments (but with real coefficients): you can save
  almost a factor of two for high degrees. It is so complicated that
  it is basically only usable via code generation, so it would be
  especially nice to modify the @horner macro to switch to this for
  complex arguments.
\end{quote}
No sooner than this was proposed, the macro was rewritten to allow for
this case giving a factor of four performance improvement on all real
polynomials evaluated at complex arguments.

\subsection{Easy and flexible parallelism}
\label{sec:easypar}


Parallel computing remains an important research topic in numerical
computing.  Parallel computing  has yet to reach the level of richness and
interactivity required for innovation that has been achieved with
sequential tools.  The  issues discussed in Section
\ref{sec:humancomputer} on the balance  between the human and the
computer become more pronounced in the parallel setting. Part of the
problem is that parallel computing means different things to different
people:
\begin{enumerate}
\item At the most basic level, one wants instruction level parallelism
  within a CPU, and expects the compiler to discover such parallelism
  in the code. In Julia, this can be achieved explicitly with the use
  of the {\tt @simd} primitive. Beyond that,
\item In order to utilize multicore and manycore CPUs on the same node, one
  wants some kind of multi-threading. Currently, we have experimental
  multi-threading support in Julia, and this will be the topic of a
  further paper. Julia currently does provide a {\tt SharedArray} data
  structure where the same array in memory can be operated on by
  multiple different Julia processes on the same node.
\item Then, there is distributed memory,
often considered the most difficult kind of parallelism.
This can mean running Julia
  on anything between half a dozen to thousands of nodes, each with
  multicore CPUs.
\end{enumerate}
In the fullness of time, there may be a unified programming model that
addresses this hierarchical nature of parallelism at different levels,
across different memory hierarchies.


Our experience with Star-P \cite{starpright} taught us a
valuable lesson.
Star-P parallelism \cite{starpstart,starpug}  included global
dense, sparse, and cell arrays that were distributed on parallel shared or distributed memory computers.  Before the evolution of the cloud as we know it today,
the user used a familiar front end (usually Matlab) as the client on a laptop or desktop, and connected seamlessly to a server (usually a large distributed computer).
Blockbuster functions from sparse and dense linear algebra, parallel FFTs,
parallel sorting, and many others were easily available and composable for the user.
In these cases Star-P called Fortran/MPI or C/MPI.
Star-P also allowed a kind of parallel for loop that worked on rows, planes or hyperplanes
of an array.  In these cases Star-P used copies of the client language on the backend,
usually Matlab, octave, python, or R.

Our experience taught us that while we were able to get a useful
parallel computing system this way, bolting parallelism onto an
existing language that was not designed for performance or parallelism
is difficult at best, and impossible at worst.  One of our (not so
secret) motivations to build Julia was to have the right language for
parallel computing.

Julia provides many facilities for parallelism, which are described in
detail in the Julia
manual\footnote{\url{http://docs.julialang.org/en/latest/manual/parallel-computing/}}. Distributed
memory programming in Julia is built on two primtives - {\it remote
  calls} that execute a function on a remote processor and {\it remote
  references} that are returned by the remote processor to the
caller. These primitives are implemented completely within Julia. On
top of these, Julia provides a distributed array data structure, a
{\tt pmap} implementation, and a way to parallelize independent
iterations of a loop with the {\tt @parallel} macro - all of which can
parallelize code in distributed memory. These ideas are exploratory in
nature, and will certainly evolve. We only discuss them here to
emphasize that well-designed programming language abstractions and
primitives allow one to express and implement parallelism completely
within the language, and explore a number of different parallel
programming models with ease. We hope to have a detailed discussion on
Juila's approach to parallelism in a future paper.

We proceed with one example that demonstrates {\tt @parallel} at work,
and how one can impulsively grab a large number of processors and
explore their problem space quickly.

\ja
\verb& &\sh{ @everywhere} \verb&begin                       & \sh {\# define on every processor} \\
\verb& function stochastic(&$\beta$\verb&=2,n=200)  &\\
\verb&   h=n^-(1/3)  & \\
\verb&   x=0:h:10  & \\
\verb&   N=length(x) & \\
\verb&   d=(-2/h^2 .-x) +  2sqrt&(h*$\beta$\verb&)*randn(N)& \sh{ \# diagonal} \\
\verb&   e=ones(N-1)/h^2                       &                   \sh {\# subdiagonal} \\
\verb&   eigvals(SymTridiagonal(d,e))[N]    & \sh   { \# smallest negative eigenvalue} \\
\verb&end&  \\
\verb&end&
\vspace{-.1in}
\end{jinput}

\ja
\vspace{-.1in}
\begin{verbatim}
t = 10000
\end{verbatim}
\vspace{-0.07in}
\verb&for &$\beta$\verb&=[1,2,4,10,20]& \\
\vspace{-0.07in}
\verb& hist([stochastic(&$\beta$\verb&) for i=1:t], -4:.01:1)[2]&
\begin{verbatim}
  plot(midpoints(-4:.01:1),z/sum(z)/.01)
end
\end{verbatim}
 \includegraphics[width=2.8in]{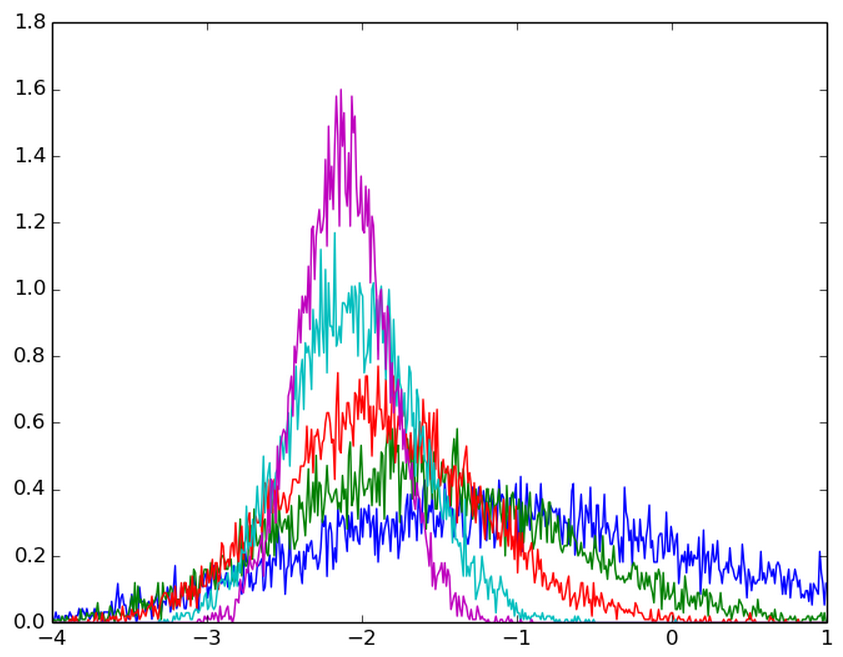}
\end{jinput}

Suppose we wish to perform a complicated histogram
in parallel. We use an example from Random Matrix Theory,
(but it could easily have been from finance),
the computation of the scaled largest eigenvalue in magnitude
of the so called stochastic Airy operator \cite{edelmansutton}
$$\frac{d^2}{dx^2}-x + \frac{1}{2\sqrt{\beta}} dW.$$

This is just the usual finite difference  discretization of
$\frac{d^2}{dx^2}-x $ with a ``noisy" diagonal.

We illustrate an example of the famous
Tracy-Widom law being simulated with Monte Carlo experiments for
different values of the inverse temperature parameter $\beta$.  The code on 1 processor is fuzzy and
unfocused, as compared to the same simulation on 1024 processors,
which is sharp and focused, and runs in exactly the same wall clock time as the sequential run. It is this ability of being able to perform scientific computation at the speed of thought conveniently without the traditional fuss associated with parallel computing, that we believe will make a new era of scientific discovery possible.

\ja
\vspace{-.1in}
\sh{\# Readily adding 1024 processors sharpens the Monte Carlo simulation in}\\ \sh{\# the same time} \\
addprocs(1024)
\vspace{-.1in}
\end{jinput}

\ja
\vspace{-.1in}
\begin{verbatim}
t = 10000
\end{verbatim}
\vspace{-0.07in}
\verb&for &$\beta$\verb&=[1,2,4,10,20]& \\
\hspace*{.07in}  \sh{z = {@parallel (+)} for p=1:nprocs()} \\
\vspace{-0.07in}
  \verb&    hist([stochastic(&$\beta$\verb&) for i=1:t], -4:.01:1)[2]& \\[.06in]
  \vspace{0.07in}
 \hspace*{.07in}   \sh {end}
\vspace{-0.12in}
\begin{verbatim}
  plot(midpoints(-4:.01:1),z/sum(z)/.01)
  end
end
\end{verbatim}
 \includegraphics[width=2.8in]{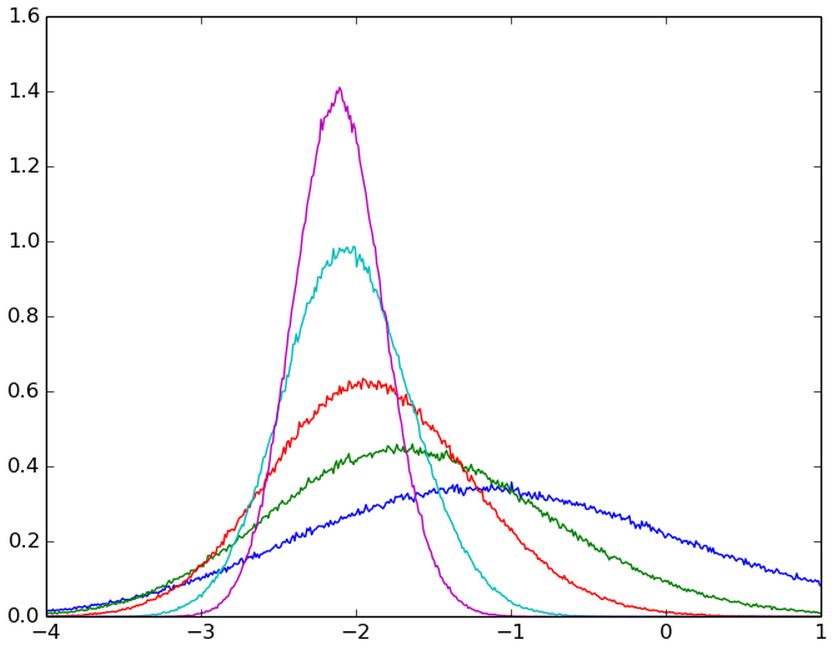}
\end{jinput}

\subsection{Performance Recap}

In the early days of high level numerical computing languages,  the thinking was that
the performance of the high level language did not matter  so long
as most of the time was spent inside the  numerical libraries.
These libraries consisted  of  blockbuster algorithms that would be highly tuned,
making efficient use of computer memory, cache, and low level instructions.

What the world learned was that
only a few codes were spending a majority of their time in the blockbusters.
Real codes were being caught by interpreter overheads, stemming from processing
more aspects of a program at run time than are strictly necessary.

As we explored in Section \ref{sec:types}, one of the hindrances of completing this analysis is
type information.  Programming language design thus becomes an exercise in
balancing incentives to the programmer to provide type
information and the ability of the computer to infer type information. Vectorization
is one such incentive system.  Existing numerical computing languages would have
us believe that this is the only system, or even if there were others, that somehow
this was the best system.

Vectorization at the software level can be elegant for some problems.
There are many matrix computation problems that look beautiful vectorized.
These programs should be vectorized. Other programs
require heroics and skill to vectorize sometimes producing unreadable code all in the name
of performance.  These are the ones that we object to vectorizing.  Still other programs
can not be vectorized very well even with heroics.
The Julia message is to vectorize when it is natural, producing nice code.  Do not vectorize
in the name of speed.

Some users believe that vectorization is required to make use of special hardware
capabilities such as SIMD instructions, multithreading, GPU units,
and other forms of parallelism.
This is not strictly true, as compilers are increasingly able to
apply these performance features to explicit loops.
The Julia message remains: vectorize when natural, when you  feel it is right.

\section{Conclusion and Acknowledgments}

We built Julia to meet our needs for numerical computing, and it
turns out that many others wanted exactly the same thing.  At the
time of writing, not a day goes by where we don't learn that someone
else has picked up Julia at universities and companies around the
world, in fields as diverse as engineering, mathematics, physical and
social sciences, finance, biotech, and many others. More than just a
language, Julia has become a place for programmers, physical
scientists, social scientists, computational scientists,
mathematicians, and others to pool their collective knowledge in the
form of online discussions and in the form of code. Numerical
computing is maturing and it is exciting to watch!

Julia would not have been possible without the enthusiasm and
contributions of the Julia
community\footnote{https://github.com/JuliaLang/julia/graphs/contributors}.
We thank Michael La Croix for his beautiful Julia display macros.
We are indebted
at MIT  to Jeremy Kepner, Chris Hill, Saman Amarasinghe,
Charles Leiserson, Steven Johnson and Gil Strang for their collegial
support which not only allowed for the possibility of an academic
research project to update technical computing, but made it more fun too.
The
authors gratefully acknowledge financial support from the MIT Deshpande center
for numerical innovation, the Intel Technology Science Center for Big
Data, the DARPA Xdata program, the Singapore MIT Alliance, NSF Awards
CCF-0832997 and DMS-1016125, VMWare Research, a DOE grant with
Dr. Andrew Gelman of Columbia University for petascale hierarchical
modeling, grants from Aramco oil thanks to Ali Dogru  and  Shell oil thanks
to Alon Arad,
and a Citibank grant for High Performance Banking Data Analysis, and the Gordon and Betty Moore foundation.

\bibliography{refs}{}
\bibliographystyle{plain}

\end{document}